\documentclass[11pt,a4paper]{article}
\pdfoutput=1
\usepackage{jheppub}

\usepackage{amsmath}
\usepackage{verbatim}
\usepackage{graphicx}
\usepackage{mathrsfs}
\usepackage{appendix}
\usepackage{caption}
\usepackage{float}
\usepackage{subfigure}
\usepackage{epstopdf}

\newcommand{\e}{\epsilon}

\newcommand{\be}[1]{ \begin{equation}\label{#1} }
\newcommand{\ee}{\end{equation}}
\newcommand{\bea}[1]{\begin{eqnarray}\label{#1} }
\newcommand{\eea}{\end{eqnarray}}
\newcommand{\bes}{\begin{subequations}}
	\newcommand{\ees}{\end{subequations}}

\newcommand{\p}{\partial}
\renewcommand{\a}{\alpha}
\renewcommand{\b}{\beta}

\newcommand{\refb}[1]{(\ref{#1})}

\newcommand{\non}{\nonumber}

\newcommand{\ie}{\emph{i.e.}}
\newcommand{\eg}{\emph{e.g.}}

\title{Galilean Gauge Theories from Null Reductions}

\author[a]{Arjun Bagchi,}\author[b]{Rudranil Basu,} \author[a]{Minhajul Islam,} \author[a]{Kedar S.~Kolekar,}\author[b,c]{Aditya Mehra.} \author{\\}
\affiliation[a]{Indian Institute of Technology Kanpur, Kalyanpur, Kanpur 208016. INDIA. \\}
\affiliation[b]{Department of Physics, BITS-Pilani, K K Birla Goa Campus, Zuarinagar, Goa-403726.
	INDIA\\}
\affiliation[c]{Max-Planck-Institut f{\"u}r Gravitationsphysik, Albert-Einstein-Institut, 14476, Golm, Germany.\\} 
\emailAdd{: abagchi@iitk.ac.in, rudranilb@goa.bits-pilani.ac.in, minhajul@iitk.ac.in, kedarsk@iitk.ac.in, aditya1.mehra@gmail.com}

\abstract{The procedure of null reduction provides a concrete way of constructing field theories with Galilean invariance. We use this to examine Galilean gauge theories, viz. Galilean electrodynamics and Yang-Mills theories in spacetime dimensions 3 and 4. Different non-relativistic conformal symmetries arise in these contexts: Schr{\"o}dinger symmetry in $d=3$ and Galilean conformal symmetry in $d=4$. A canonical analysis further reveals that the symmetries enhance to their infinite dimensional versions in phase space and pick up central extensions. In addition, for the Abelian theory, we discuss non-relativistic electro-magnetic duality in $d=3$ and its difference with the $d=4$ version. We also mention some quantum aspects for both Abelian and non-Abelian theories.}

\preprint{}

\begin{document}
	
\maketitle

\bigskip	
\section{Introduction}
Our present understanding of Nature rests heavily on the framework of quantum field theory. For the theory of elementary particles, naturally this comes hand in hand with relativistic physics. So the formulation of quantum field theories, specifically gauge theories, are intimately connected with Lorentz or Poincar{\'e} symmetry. For various applications, it is necessary to look beyond relativistic theories. For effective descriptions of systems where typical velocities are much lower than the speed of light, the Galilean algebra replaces the Poincar{\'e} algebra as the symmetry of interest. This happens, e.g. in many real-life condensed matter systems, (non-relativistic) hydrodynamics. For systems such as these, quantum field theoretic descriptions naturally to incorporate Galilean symmetries instead of Lorentz or Poincar{\'e} symmetry. Although this seems like an age-old problem, interesting progress in the field has been achieved only recently, coinciding with an effort in understanding the Holographic Principle away from its usual setting in relativistic Anti-de Sitter (AdS) spacetimes \cite{Son:2008ye,Balasubramanian:2008dm,Kachru:2008yh,Bagchi:2009my,2009}. In this paper, we shall be interested in the construction of Galilean gauge theories. Initial works on Galilean gauge theories date back to the construction of Galilean electrodynamics by Le Ballac and Levy-Leblond \cite{LBLL} in the 1970s. The more recent papers in this direction are \cite{Bagchi:2014ysa,Duval:2014uoa,Bagchi:2015qcw,VandenBleeken:2015rzu,Bergshoeff:2015sic,Festuccia:2016caf, Banerjee:2019axy,Mehra:2021sfx}.  

Relativistic quantum field theories are defined on fixed (pseudo-) Riemannian manifolds with a definite metric structure. When we move to Galilean field theories, the background metric on the manifold becomes degenerate as the speed of light goes to infinity and the (pseudo-) Riemannian structure is replaced by a so-called Newton-Cartan structure which is a quadruple $(\mathcal{G}, h, \tau, \Gamma)$ \cite{Duval:1993pe,Duval:2009vt}. $\mathcal{G}$ is a $(d+1)$ dimensional manifold where we can choose a set of coordinates $(t, x^i)$. $h$ is a contravariant spatial metric $h^{\mu\nu}$ and $\tau=\tau_a dx^a$ a non-vanishing one-form called the clock form. $\Gamma$ is torsion-free linear connection. A flat NC structure is given by a vanishing $\Gamma$. The Galilean algebra is defined as the Lie algebra of vector fields $\xi=\xi^a\p_a$ satisfying 
\be{gal-nc}
\mathcal{L}_\xi h^{\mu\nu} = 0, \quad  \mathcal{L}_\xi \tau_a =0, \quad \tau_ah^{ab}=0.
\ee
The algebra generated is the same as the one obtained by an In{\"o}n{\"u}-Wigner contraction of the Poincar{\'e} algebra. 

The absence of a non-degenerate metric makes the writing of an action a difficult proposition. Some of the recent literature, include some of our earlier work in this direction \cite{Bagchi:2014ysa,Bagchi:2015qcw,Bagchi:2017yvj}, thus concentrated on the aspects of symmetries arising from the equations of motion (EOM) which were arrived at by a limiting procedure starting from a relativistic quantum field theory. Particular attention was given to the conformal versions of these Galilean QFTs, or Galilean Conformal Field Theories. It was shown that Galilean gauge theories in four dimensions, viz. Galilean electrodynamics and Galilean Yang-Mills theories had EOM that exhibited invariance under the infinite dimensional Galilean Conformal Algebra. These infinite enhancements remained when one coupled the gauge theories to massless matter (scalars or fermions). We will review some of this briefly later in the paper. But our focus in this current paper is the construction of explicit actions for these gauge theories. We do this by the process of null reduction \cite{Duval:1984cj,1991,1995,Santos:2004pq} that has been previously employed for generating non-relativistic theories from relativistic theories in one higher dimension.  

In particular, we focus on Galilean gauge theories in dimensions three and four. We start with relativistic gauge theories in $d=4, 5$ and null reduce to generate actions for Galilean electrodynamics and Galilean Yang-Mills theories in $d=3,4$. For the abelian theories, this process yields results that are already present in the literature, reproducing the so-called Magnetic sector of GED in $d=4$ \cite{Bagchi:2014ysa,Festuccia:2016caf} and a known action in $d=3$ \cite{Chapman:2020vtn}. The actions of Galilean YM are again straight-forward generalisations of their abelian counterparts. Interesting results appear when we delve deeper into the symmetries of these actions and their corresponding EOM. We find that for the gauge theories non-relativistic conformal algebras, the Galilean Conformal Algebra in $d=4$ and the Schr\"odinger algebra in $d=3$ are realised at the level of the action. These are actually partially enhanced to their infinite dimensional algebras in the action itself. At the level of the EOM, there is further enhancement of symmetries. While the appearance of the infinite dimensional GCA is not unexpected given earlier works, the appearance of the Schr\"odinger-Virasoro algebra in $d=3$ is new. Further surprises await us when we look at the canonical analysis later in the paper. Here we find that the entire infinite dimensional algebras are realised as symmetries in phase space. There are also central extensions which we find. 

Along the way, we also investigate aspects of electromagnetic duality for Galilean Electromagnetism. While in $d=4$ the duality has been known to exchange electric and magnetic theories, we find that in $d=3$ the theory itself has an in-built electromagnetic duality. This is inherited from the null reduction of the higher dimensional relativistic theory. We end our analysis with a few quantum aspects of Galilean Yang-Mills theory. A detailed exposition of this will be carried out in later work. 	

\paragraph{Outline:} The paper is organised as follows. We begin in Sec~2 with a review of non-relativistic conformal symmetries, specifically focussing on the Galilean conformal algebra and then the Schr{\"o}dinger algebra first discussing the finite dimensional algebras and then going onto their infinite dimensional extensions. We briefly comment about their geometric realisations. We end the section with a quick review of the procedure of null reduction. 

In Sec.~3, we address Abelian Galilean gauge theories. We begin with some review material about Galilean electrodynamics in $d=4$ and then proceed to find actions by null reductions in $d=4$ and $d=3$. Symmetries arising from the actions and the corresponding EOM are discussed. While most of this is review material, we discover a realisation of the infinite dimensional Schr{\"o}dinger-Virasoro algebra in $d=3$. We then discuss electromagnetic duality for the $d=3$ Galilean theory and its differences in the $d=4$ case. Much of the $d=3$ analysis is based on null reductions. 	
	
Sec.~4 addresses non-Abelian theories. The section is structured similar to the Abelian one, but we provide more details of the symmetry analyses of the actions and EOM. Unlike the Abelian case, we find that our EOMs don't reduce to earlier work in $d=4$ when some fields are turned off. Interestingly, though we again find an enhancement of symmetries in $d=4$, as well as in $d=3$. 

We perform a canonical analysis of symmetries in Sec.~5 and find that the infinite dimensional Virasoro sub-algebras are non-trivially realised in phase space for the Yang-Mills theories in both dimensions. In addition, there are state-dependent central extensions to the Virasoro algebra. In Sec.~6, we briefly discuss quantum mechanical aspects and write down the Feynman rules for Galilean Yang-Mills theories. We conclude with a summary of our results and a list of future directions in Sec.~7. An appendix contains some aspects of GCA representation theory.

\bigskip

\section{Non-relativistic Conformal Symmetries}
Different versions of conformal symmetry have been discussed in the non-relativistic setting \cite{Taylor:2008tg}. We will focus our attention on two of these, viz. Galilean Conformal symmetry \cite{Bagchi:2009my} and Schr\"odinger symmetry \cite{Hagen:1972pd,Niederer:1972zz,Henkel:1993sg,Son:2008ye}. Galilean conformal symmetry is arrived at by an Inonu-Wigner contraction of relativistic conformal symmetry, much the same way as one derives the Galilean algebra from the Poincar{\'e} algebra. This is the symmetry for a massless or gapless non-relativistic system. Schr\"odinger symmetry on the other hand is the group of symmetries of the free Schr\"odinger equation. 

In order to set notation, let us remind ourselves of the relativistic conformal generators in $d$ dimensional Minkowski spacetime. The conformal group is constructed from translations [$P_{\mu}$], Lorentz transformations $[J_{\mu\nu}]$, dilatation [$D$], and special conformal transformations (SCT) [$K_{\mu}$]. In vector field notation, they are given by
	\begin{eqnarray}\label{CO}
	P_{\mu}=\partial_{\mu}, \quad J_{\mu\nu}=(x_{\mu}\partial_{\nu}-x_{\nu}\partial_{\mu}), \quad D=-x^{\mu}\partial_{\mu}, \quad K_{\mu}=-(2x_{\mu}x^{\nu}\partial_{\nu}-x^{\nu}x_{\nu}\partial_{\mu}),
	\end{eqnarray}
where $\mu=(0,i=1,...,d-1)$. The commutation relations satisfied by these generators form the conformal algebra, which is isomorphic to $so(d,2)$.

\subsection{Galilean Conformal Symmetries}
	
As stated earlier, the Galilean Conformal Algebra is obtained from the relativistic conformal algebra through an In\"{o}n\"{u}-Wigner contraction. We will perform the non-relativistic limit at the spacetime level and in units where $c=1$. In this limit, the velocities are small as compared to the speed of light. The systematic procedure to take the limit is as follows
	\begin{eqnarray}\label{GL}
	x_{i}\rightarrow \epsilon x_{i}, \,t\rightarrow t \quad \text{along with}~~\epsilon \rightarrow 0.
	\end{eqnarray}
The scaling in \eqref{GL} is equivalent to $c\rightarrow \infty$ scaling. Contracting \eqref{CO} in this way we get
	\bes{}\label{GCA operator}
	\begin{eqnarray}&&
	H= -\partial_{t}, \quad  D=-(t\partial_{t}+x^{i}\partial_{i}), \quad K=-(t^{2}\partial_{t}+2tx^{i}\partial_{i}),\\&&
	P_{i}=\partial_{i},\quad B_{i}=t\partial_{i},\quad K_{i}=t^{2}\partial_{i},\\&&
	J_{ij}=(x_{i}\partial_{j}-x_{j}\partial_{i}).
	\end{eqnarray}\ees
	where $(H,D,K)\rightarrow$ \{Hamiltonian, dilatation and temporal SCT\} and $(P_i,B_i,K_i)\rightarrow$ \{translation, boost and spatial SCT\}. Finally, $J_{ij}$ are the generators of spatial rotations.
	The algebra of these generators are given by
	\begin{eqnarray}\label{alg}
	&&[J_{ij},J_{kl}]=(\delta_{ik}J_{jl}-\delta_{jk}J_{il}-\delta_{il}J_{jk}+\delta_{jl}J_{ik}), \quad [H,B_i]=P_i,\non\\\non&&
	[P_i,B_j]=0, \quad [J_{ij},X_k]=(\delta_{ik}X_j-\delta_{jk}X_i), \quad [B_i,B_j]=0,\\&&
	[D,H]=H,\quad [D,P_i]=P_i, \quad [D,K]=-K, \quad	[D,K_i]=-K_i,\non \\ &&[K,H]=2D, \quad [K,P_i]=2B_i, \quad [B_i,K]=-K_i, \non\\ &&
	[B_i,K_j]=0, \quad [K_i,P_j]=0, \quad [K_i, H]=2B_i,
	\end{eqnarray}
	where $X_i=(P_i, B_i, K_i)$. This is a finite dimensional Galilean conformal algebra (fGCA). The generators can also be written in a suggestive form for $n=0,\pm 1$ given as
	\bes{}
	\begin{eqnarray}&&
	L^{(n)}=-t^{n+1}\partial_{t}-(n+1)t^{n}x^{i}\partial_{i}, \quad (L^{(-1,0,1)} \rightarrow H,D,K)\\&&
	M_{i}^{(n)}=t^{n+1}\partial_{i}, \quad\quad (M^{(-1,0,1)}_i \rightarrow P_{i}, B_{i}, K_{i})
	\end{eqnarray}\ees
	The fGCA can be written down as
	\begin{eqnarray}&&
	\big[L^{(n)},L^{(m)}\big]=(n-m)L^{(n+m)},~ \big[L^{(n)},M_{i}^{(m)}\big]=(n-m)M_{i}^{(n+m)},~ \big[M^{(n)}_{i},M^{(m)}_{j}\big]=0,\non\\&&
	\big[L^{(n)},J_{ij}\big]=0,~ \big[J_{ij},M^{(n)}_{k}\big]=M^{(n)}_{j}\delta_{ik}-M^{(n)}_{i}\delta_{jk}.
	\end{eqnarray}
	One sees from above that the algebra continues to hold for any integer value of $n$. Surprisingly the algebra is extended to an infinite-dimensional algebra which we name I-GCA or simply GCA from now on. The algebra above can admit central terms in the usual Virasoro subalgebra. There is a further central extension which the $[L, M]$ commutator gets in the special case of $d=2$. In the systems we study later in the paper, we will find that in their realisation in phase space, the symmetries develop a Virasoro central extension. 
	
The action of infinite extension of GCA on the fields at general space-time points is given by \cite{Bagchi:2009ca,Bagchi:2014ysa,Bagchi:2017yvj}: 
	\bes{}\label{gca rep}
	\begin{eqnarray}&&\label{gcaln}
	\hspace{-1.5cm}\big[L^{(n)},\varphi(t,x)\big]=\big(t^{n+1}\partial_{t}+(n+1)t^{n}x^{k}\p_{k}+(n+1)t^{n}\Delta\big)\varphi(t,x)\non\\&&
	\hspace{2.7cm}-n(n+1)t^{n-1}x^{k}U\big[M_{k}^{(0)},\varphi(0,0)\big]U^{-1},\\&&\label{gcamn}
	\hspace{-1.5cm}\big[M^{(n)}_{k},\varphi(t,x)\big]=-t^{n+1}\partial_{k}\varphi(t,x)
	+(n+1)t^{n}U\big[M_{k}^{(0)},\varphi(0,0)\big]U^{-1}.
	\end{eqnarray}\ees
	In the above, $\varphi(t,x)$ represents a field of generic spin. For the purpose of this paper, we shall only be interested in scalars and gauge fields and hence the transformation laws read:
	\begin{equation}\label{boot}
	\big[M_{k}^{(0)},\varphi_A(0,0)\big]=a\phi_k \delta_{A0}+s\delta_{At}a_{k} + r\delta_{Ai}\phi \delta_{ik},
	\end{equation}
	where the index $i$ is not summed over and $\varphi_0(0,0)=\phi$, $\varphi_t(0,0)=a_t$, $\varphi_i(0,0)= a_i$ $(\phi \rightarrow \text{scalar field} ,\{a_t,a_i\}\rightarrow \text{gauge fields})$. The values of constants $(a,s,r)$ are determined by demanding inputs from the dynamics. We will clarify later in the paper how the values of these constants get fixed for the theory under consideration (see \eg\ Sec[\ref{sec:GYM-symmetries}]).
We will use \eqref{gca rep} to check the symmetries of the system we are dealing with in this paper. For the detailed analysis of the actions of GCA generators on fields, the reader is directed to appendix \ref{app:representation}.

\subsection{Schr\"{o}dinger Symmetries}

We now revisit another non-relativistic algebra that has less symmetries in comparison to GCA. The Schr\"{o}dinger group \cite{Hagen:1972pd,Niederer:1972zz},  $Sch(d)$ is a symmetry group of the free Schr\"{o}dinger equation in $d$ dimensions. This symmetry has been discussed in the context of fermions at unitarity \cite{Nishida:2007pj}. Some recent literature on the Schr\"{o}dinger algebra includes \cite{Hellerman:2021qzz,Karananas:2021bqw,Pellizzani:2021hzx,Hellerman:2020eff,Kravec:2018qnu, Favrod:2018xov,Pal:2018idc}.

The Schr\"{o}dinger algebra is generated by the generators of centrally extended Galilean algebra with two additional generators; the dilatation $\tilde{D}$ and a special conformal transformation $\tilde{K}$. These generators are given by
	\begin{eqnarray}
	&& P_{i}=\p_{i}, \quad H=\p_t, \quad G_{i}=t\p_{i}+x_{i}M, \quad J_{ij}=(x_{i}\p_{j}-x_{j}\p_{i}),  \nonumber \\
	&& \tilde{D}=2t\p_t+x^i\p_{i}, \quad \tilde{K}={2}t^{2}\p_{t}+2tx^{i}\p_{i}+x^{2}M.
	\end{eqnarray}
	The non-zero commutations relations of the Schr\"{o}dinger algebra are 
	\begin{eqnarray}
	&& [H,\tilde{D}]=2H,\quad [P_{i},\tilde{D}]=P_{i},\quad [\tilde{D},G_{i}]=G_{i},\quad [\tilde{D},\tilde{K}]=2\tilde{K}, \nonumber \\
	&& [P_{i},G_{j}]=-\delta_{ij}M,\quad [H,\tilde{K}]=\tilde{D},\quad [H,G_{i}]=-P_{i},\quad [G_{i},\tilde{K}]=-G_{i}, \nonumber \\
	&& [J_{ij},P_{k}]=\delta_{ik}P_{j}-\delta_{jk}P_{i},\quad [J_{ij},G_{k}]=\delta_{ik}G_{j}-\delta_{jk}G_{i}, \nonumber \\
	&& [J_{ij},J_{kl}]=(\delta_{ik}J_{jl}+\delta_{jl}J_{ik}-\delta_{il}J_{jk}-\delta_{jk}J_{il}).
	\end{eqnarray}
Here $G_i$ is the Schr\"{o}dinger boost, and $M$ is a central element, that can be viewed as the mass or the particle number of the system. The number of generators of the group, including the central element $M$, in $d=4$ is 13, which is to be contrasted with the finite GCA, which has 15. The difference comes about in the number of special conformal generators, which is a vector-worth in the GCA compared to a single one in the Schr{\"o}dinger algebra.

We now briefly discuss the representation of the Schr\"{o}dinger algebra following \cite{Nishida:2007pj}. A primary operator $\varphi(0,0)$ at $(t=0,\ x^i=0)$ of scaling dimension $\Delta_{\varphi}$ is defined by
	\begin{equation}
	[\tilde{D},\varphi(0,0)] = \Delta_{\varphi}\varphi(0,0); \quad [G_i, \varphi(0,0)] = 0, \quad [\tilde{K},\varphi(0,0)] = 0.
	\end{equation}
	Then the tower of operators built by repeated action of $P_i$ and $H$ on a primary operator forms an irreducible representation of the Schr\"{o}dinger algebra. In this way, we can see that all local operators can be divided into irreducible representations built on different primary operators. For a primary operator at an arbitrary spacetime point $(t,x^i)$, the generators $G_i$ and $\tilde{K}$ act as
	\bea{}\label{gk}&&
	\big[G_{i},\varphi(t,x)\big]=(t\p_{i}+m_{\varphi}x_{i})\varphi(t,x),\non\\&& \big[\tilde{K},\varphi(t,x)\big]=[(2t^{2}\p_{t}+2tx^{i}\p_{i}+t\Delta_{\varphi})+x^{2}m_{\varphi}]\varphi(t,x),
	\eea
	where $m_{\varphi}$ is the eigenvalue of $M$ on $\varphi$. For any local operator at an arbitrary spacetime point, the generators $P_i$, $H$ and $\tilde{D}$ act as
	\bea{}\label{phd}&&
	\big[P_{i},\varphi(t,x)\big]=\p_{i}\varphi(t,x),~ \big[H,\varphi(t,x)\big]=\p_{t}\varphi(t,x),\non\\&& \big[\tilde{D},\varphi(t,x)\big]=(2t\p_{t}+x^{i}\p_{i}+\Delta_{\varphi})\varphi(t,x).
	\eea
	The boost and dilatation generators act on the fields of our interest $\varphi=(\phi \rightarrow \text{scalar field} ,\\
		\{a_t,a_i\}\rightarrow \text{gauge fields})$ as
	\bes{}
	\bea{}\label{bootsa}&&
	\delta_{G_{k}} \phi = -t\partial_{k}\phi,~\delta_{G_{k}} a_{t}= -(t\partial_{k}a_{t}+a_{k}),~\delta_{G_{k}} a_{i}=-(t\partial_{k}a_{i}-\delta_{ik}\phi),\\&&\label{stsa}
	\delta_{\tilde{D}} \varphi=(2t\partial_{t}+x^{k}\partial_{k}+\Delta_{\varphi})\varphi.
	\eea\ees
	The generator of the special conformal transformation $\tilde{K}$ acts on $\varphi$ as
	\bes{}\label{sctsa}
	\bea{}&&
	\delta_{{\tilde{K}}} a_{t}=(2t^{2}\partial_{t}a_{t}+2tx^{k}\partial_{k}a_{t}+4ta_{t}+2x^{k}a_{k}),\\&&
	\delta_{\tilde{K}} a_{i}=(2t^{2}\partial_{t}a_{i}+2tx^{k}\partial_{k}a_{i}+2ta_{i}-2x_{i}\phi),\\&&
	\delta_{\tilde{K}} \phi = (2t^{2}\partial_{t}\phi+2tx^{k}\partial_{k}\phi).
	\eea	\ees
Since we would be interested in gauge fields in the bulk of the paper, we would be dealing with massless systems and hence will restrict ourselves to the zero eigenvalue subsector of $M$ $:m_{\varphi}=0$.
	
Although the Schr\"{o}dinger algebra as reviewed above is rather well known, somewhat less studied is the interesting observation that even this algebra admits an infinite extension \cite{Henkel:1993sg, Alishahiha:2009nm}. To get this infinite extension, we first rewrite the generators as
	\begin{eqnarray}\label{zyt}
	&& Z^{(n)}=2t^{n+1}\p_{t}+(n+1)t^{n}x^{i}\p_{i}+\frac{1}{2}n(n+1)t^{n-1}x^{2}M, \nonumber \\
	&& Y_{i}^{(m)}=t^{m+\frac{1}{2}}\p_{i}+\big(m+\frac{1}{2}\big)t^{m-\frac{1}{2}}x_{i}M, \nonumber \\
	&& T^{(n)}=t^{n}M,
	\end{eqnarray}
	where $n=(-1,0,1)$, $m=(-\frac{1}{2},\frac{1}{2})$ and $(Z^{(-1,0,1)} \rightarrow H,\tilde{D},\tilde{K})$, $(Y^{(-\frac{1}{2},\frac{1}{2})}_{i} \rightarrow P_{i}, G_{i})$. To distinguish from the GCA case, we use the notation $Z^{(n)}$ instead of the usual $L^{(n)}$ notation for the Schr\"{o}dinger case. In terms of these redefined generators the Schr\"{o}dinger algebra becomes
\begin{eqnarray}\label{schrod}
	&& [Z^{(n)},Z^{(m)}]=(n-m)Z^{(n+m)},\quad [Y_{i}^{(n)},Y_{j}^{(m)}]=(n-m)\delta_{ij}T^{(n+m)}, \nonumber \\
	&& [Z^{(n)},Y_{i}^{(m)}]=(\frac{n}{2}-m)Y_{i}^{(n+m)}, \quad [Z^{(n)},T^{(m)}]=-mT^{(n+m)}.
\end{eqnarray}
We see that the above algebra is satisfied if we extend the values of $n$, $m$ to $n\in \mathbb{Z}$, $m\in \mathbb{Z}+\frac{1}{2}$, thus giving us an infinite extension of the Schr\"{o}dinger algebra, which in literature is referred to as the Schr\"{o}dinger-Virasoro algebra. The Virasoro sub-algebra again admits a central extension, which we will encounter in our analysis later in the paper. The action of infinite generators $Z^{(n)}$ on our fields of interest $\varphi = (\phi,a_t,a_i)$ can be written down as
	\begin{eqnarray}\label{3d L transformation abelian1}&&
	\delta_{Z^{(n)}}\phi=2t^{n+1}\p_t \phi+(n+1)t^{n}(x^{k}\partial_{k}+\Delta_{\phi})\phi, \nonumber \\&&
	\delta_{Z^{(n)}}a_{i}=2t^{n+1}\p_t a_{i}+(n+1)t^{n}(x^{k}\partial_{k}+\Delta_{a_i})a_{i}-n(n+1)t^{n-1}x_{i}\phi, \nonumber \\&&
	\delta_{Z^{(n)}}a_{t}=2t^{n+1}\p_t {a}_{t}+(n+1)t^{n}(x^{k}\partial_{k}+\Delta_{a_t})a_{t}+n(n+1)t^{n-1}x^{k}a_{k}.
	\end{eqnarray}
	Similarly the action of $Y^{(n)}_{k}$ generators on the fields $\varphi$ is
	\begin{eqnarray}\label{3d M trans abelian1}&&
	\delta_{Y^{(m)}_{k}}\phi=t^{m+\frac{1}{2}}\, \partial_{k}\phi,~
	\delta_{Y^{(m)}_{k}}a_{i}=t^{m+\frac{1}{2}}\partial_{k}a_{i}-\big(m+\frac{1}{2}\big)t^{m-\frac{1}{2}}\phi,\non\\&&
	\delta_{Y^{(m)}_{k}}a_{t}=t^{m+\frac{1}{2}}\partial_{k}a_{t}+\big(m+\frac{1}{2}\big)t^{m-\frac{1}{2}}a_{k}.
	\end{eqnarray}
	These transformations above for the Abelian fields $(\phi,a_t,a_i)$ (and their generalizations to non-Abelian fields) will be useful to show the invariance of Galilean versions of electrodynamics (and Yang-Mills) under Schr\"{o}dinger algebra in $3$ dimensions.

\subsection{Geometrical realization of non-relativistic conformal symmetries}
In the introduction, we described the Galilean algebra as the isometry algebra of a flat Newton-Cartan manifold \refb{gal-nc}. The non-relativistic conformal algebras that we have discussed above from an algebraic perspective can also be look at geometrically. In terms of the previously discussed Newton-Cartan structures, these symmetries arise as conformal isometries \cite{Duval:2009vt,Duval:2014lpa}. 

\medskip

Generically, the infinite dimensional conformal isometries of the flat NC spacetime are generated by vector fields $\xi^a$ which satisfy the non-relativistic conformal Killing equations
\be{}
\mathcal{L}_\xi h^{\mu\nu} = \lambda h^{\mu\nu}, \quad  \mathcal{L}_\xi \tau_a = \mu \tau_a; \quad \lambda + N\mu = 0.
\ee
Solving these equations, the vectors fields $\xi^a$ span the \emph{conformal Galilei algebra of level $N$}, denoted by $\mathfrak{cgal}_N(\mathcal{G},h,\tau)$. The level $N$ of the algebra is related to the dynamical exponent $z$, which characterizes the unequal scaling of space and time under dilatation, as $z=\frac{2}{N}$. 

\medskip

For $N=2$, \ie\ $z=1$ the algebra $\mathfrak{cgal}_2(\mathcal{G},h,\tau)=\mathfrak{gca}(d+1)$ is the infinite dimensional Galilean conformal algebra. For $N=1$, \ie\ $z=2$ we get the infinite dimensional Schr\"{o}dinger-Virasoro algebra $\mathfrak{cgal}_1(\mathcal{G},h,\tau)=\mathbf{sv}(d+1)$. 

Further demanding that the non-relativistic conformal transformations preserve the form of the geodesic equation \ie\ the projective structures associated with the flat connection $\Gamma=0$, the algebra reduces to finite dimensional Schr\"{o}dinger algebra $\mathfrak{cgal}_1(\mathcal{G},h,\tau,\Gamma)=\mathfrak{sch}(d+1)$ for $z=2$, and finite dimensional GCA for $z=1$. We refer the reader to \cite{Duval:2009vt} for further detials.

	\subsection{Null Reduction}
Before moving onto the main body of the paper, let us briefly review the process of null reduction which will be our primary tool in constructing non-relativistic actions. Null reduction is a method to obtain a $d$-dimensional Galilean invariant theory from a $(d+1)$-dimensional relativistic theory \cite{Duval:1984cj,1991,1995,Santos:2004pq}. Below we elucidate this by applying it to a scalar field theory.
	
To begin with, consider the metric of a $(d+1)$-dimensional Minkowski spacetime
	\begin{eqnarray}\label{ofc}
	ds^2=\eta_{\mathcal{B}\mathcal{C}}dx^{\mathcal{B}}dx^{\mathcal{C}}=-(dx^{0})^{2}+(dx^{1})^{2} + \delta_{AB}dx^A dx^B,
	\end{eqnarray}
	where $\mathcal{B},\mathcal{C}=(0, 1, ..., d)$ and $A,B=(2,\dots,d)$. We will make a change of coordinates to the usual light-cone coordinates given below
	\begin{eqnarray}\label{ofc1}
	u=\frac{1}{\sqrt{2}}(x^{0}-x^{1}),\quad t=\frac{1}{\sqrt{2}}(x^{0}+x^{1}).
	\end{eqnarray}
	In lightcone coordinates, the metric becomes
	\begin{eqnarray}\label{ofc2}
	ds^{2}=\eta_{\tilde{\mu}\tilde{\nu}}dx^{\tilde{\mu}}dx^{\tilde{\nu}}=2du\,dt + \delta_{AB}dx^A dx^B,
	\end{eqnarray}
	where $\tilde{\mu}=(u,t,2,3,...)$. We will now look how to implement this technique on a field theory. We will first start with an action of Lorentz invariant theory on the coordinate system \eqref{ofc}. We then reduce the action of that theory along the $u$ null direction of the coordinate system \eqref{ofc2}. Here, the components of $(d+1)$-dimensional generic fields reduce to different fields in $d$-dimensional Galilean invariant theory, \eg\ a gauge field reduces as $\mathcal{A}_{\tilde{\mu}} = (\mathcal{A}_{u}, \mathcal{A}_t, \mathcal{A}_i)$. The fields $\mathcal{A}$ are taken to be independent of the coordinate $u$. The $t$ coordinate becomes time in $d$-dimensional Galilean theory and $(A,B)$ becomes the spatial coordinates $ i,j=(1,2,3,...)$.
	
	Let us discuss this technique through an example of a scalar field theory. Consider a relativistic scalar field theory in $5$-dimensional Minkowski spacetime, described by the action
	\bea{}\label{sfs}\mathcal{S}=\int dt\,d^{4}x \big(\frac{1}{2}\eta^{\mathcal{B}\mathcal{C}}\partial_{\mathcal{B}}\phi\partial_{\mathcal{C}}\phi\big)=\int dt\,d^{4}x\big(-\frac{1}{2}(\p_0 \phi)^2 + \frac{1}{2}(\p_1\phi)^2 +\frac{1}{2}(\p_i\phi)^2\big),\eea
	where $\mathcal{B}=[0,1,i=(2,3,4)]$. We will now express \eqref{sfs} in coordinate system \eqref{ofc2} so that we can perform a null reduction. The action becomes
	\begin{eqnarray}\label{sfsnr}&&
	\mathcal{S}=\int d^{5}x\big(\frac{1}{2}\eta^{\tilde{\mu}\tilde{\nu}}\partial_{\tilde{\mu}}\phi\partial_{\tilde{\nu}}\phi\big)\non\\&&
	\hspace{.82cm}=\int dt\,du\,d^{3}x\big(\frac{1}{2}\eta^{tu}\partial_{t}\phi\partial_{u}\phi+\frac{1}{2}\eta^{ut}\partial_{u}\phi\partial_{t}\phi+\frac{1}{2}\eta^{ij}\partial_{i}\phi\partial_{j}\phi\big).
	\end{eqnarray}
	With $\phi(t,x^i)$ independent of the null coordinate $u$, we perform the null reduction on \eqref{sfsnr} along $u$-direction and get the action as
	\begin{equation}\label{gsfs}
	\mathcal{S}_{G}=\int dt\,d^{3}x \big(\frac{1}{2}\eta^{ij}\partial_{i}\Phi\partial_{j}\Phi\big)
	\end{equation}
	where the subscript $G$ denotes Galilean and  $\phi(t,x^i)$ has reduced to $\Phi(t,x^i)$. This resultant action \eqref{gsfs} is invariant under GCA in $d=4$ dimensions.

\newpage
	
	
	\section{Abelian Galilean Gauge theory}
	In this section, we will address Galilean electrodynamics in dimensions $d=4$ and then $d=3$. We will construct actions by null reducing relativistic electrodynamics in one higher dimension and investigate the symmetries associate with the action and the corresponding EOM. Much of this is review material, although the discovery of the (partial) realisation of the infinite dimensional Schr{\"o}dinger Virasoro in the $d=3$ case is new. We then address the question of electromagnetic duality in these theories, focussing on the $d=3$ case and point out differences to the one found earlier in the $d=4$ case. 
	
	\subsection{Galilean electrodynamics in $d=4$}\label{4d-GED-review}
	Le Bellac and L\'{e}vy-Leblond first introduced Galilean electrodynamics (GED) in \cite{LBLL} and wrote it down in the language of electric ($\vec{E}$) and magnetic ($\vec{B}$) fields. In \citep{Bagchi:2014ysa}, we (a subset of the current authors) discussed the Galilean electrodynamics in terms of potential formulation, scaling the scalar and vector potentials in addition to the scaling of coordinates \eqref{GL}. The non-relativistic scaling of the $4$-vector potential was considered in two different ways:
	\bea{sca} a_t \rightarrow a_t, \, \, a_i \rightarrow \epsilon a_i ~~\text{and}~~ a_t \rightarrow \epsilon a_t, \,  \, a_i \rightarrow  a_i.\eea 
	The first limit is known as the Electric limit whereas the second limit is called as Magnetic limit. The EOM of these two sectors (in absence of sources) are as follows:
	\bes{}\label{dosa}
	\bea{}
	\hspace{-2cm}\mbox{{\bf{Electric limit}}:}\qquad &&\partial^i \partial_i a_t = 0, \quad \partial ^j \partial_j a_i - \partial _i \partial_j a^j + \partial_t \partial _i a_t= 0; \label{Eeom} \\
	\mbox{{\bf{Magnetic limit}}:} \qquad &&\partial ^j \partial_j a_{i}-\partial_{i}\partial_{j}a^{j} =0, \quad \partial^i \partial_i a_{t}-\partial_{i}\partial_{t}a^{i}= 0. \label{Meom}
	\eea \ees
	These equations can also be written in terms of $\vec{E}$ and $\vec{B}$. They become
	\bes{}\label{dem}
	\begin{eqnarray}
	\mbox{{\bf{Electric limit}}:}\quad&&	\vec{\nabla}\cdot\vec{E}=0, \quad \vec{\nabla}\cdot\vec{B}=0, \quad \vec{\nabla}\times\vec{E}=0,\quad \vec{\nabla}\times\vec{B}=\frac{\p\vec{E}}{\p t}\\
	\mbox{{\bf{Magnetic limit}}:}\quad&&	\vec{\nabla}\cdot\vec{E}=0, \quad \vec{\nabla}\cdot\vec{B}=0, \quad \vec{\nabla}\times\vec{E}=-\frac{\p\vec{B}}{\p t}, \quad \vec{\nabla}\times\vec{B}=0.
	\end{eqnarray}\ees
	By looking at \eqref{dem}, we notice that under the exchange of electric and magnetic fields, i.e. under the following transformation
	\begin{eqnarray}
	\vec{E} \longrightarrow \vec{B}, \quad \vec{B} \longrightarrow -\vec{E}, \label{4d-GED-em-duality}
	\end{eqnarray}
	the EOM in the electric and the magnetic sectors get interchanged. This is referred to as the electric-magnetic duality between ``electric" and ``magnetic" sectors of Galilean electrodynamics \cite{Duval:2014uoa}.
	
	\medskip
	
	\noindent We will now look at the invariance of EOM \eqref{dosa} of Galilean electrodynamics under infinite-dimensional Galilean conformal symmetry. For that, we have to use the details of representation theory (look at Appendix \ref{app:representation}). The transformation of equations of Electric limit under the generators $(L^{(n)}, M^{(n)}_{k})$ of GCA are given by
	\bea{}\label{infchecka}&&
	[ L^{(n)},\partial^i  \partial_i a_t ] = 0,~[ L^{(n)},\partial^j \partial_j a_i  -\partial_i  \partial^j  a_j  +  \partial_t \partial_ i a_t ] =- \frac{1}{2}\, n(n+1)\,t^{n-1} (d -4) \partial_i a_t ,\non
	\\&& [ M^{(n)}_k ,\partial^i  \partial_i a_t ] = 0,~[ M^{(n)}_k,\partial^j \partial_j a_i  -\partial_i  \partial^j  a_j  +  \partial_t \partial_ i a_t ] = 0.~~~~~~~~~~~~~~~~~~~~~~~~~~~~~~~~~~~~
	\eea
	We see that for $d=4$, the equations are invariant under the full infinite dimensional GCA. 
	Similarly for Magnetic case, we have
	\bea{}\label{infcheckma}&&
	[ L^{(n)}, \partial^{j}\partial_j a_{i}-\partial_{i}\partial^{j}a_{j}]=0,~
	[ L^{(n)},\partial^{i}\partial_{i} a_{t}-\partial_{t}\partial^{i}a_{i}]=-\frac{1}{2}n(n+1)(d-4)t^{n-1}\partial^{i}a_i ,\non\\&&
	[ M^{(n)}_{k},\partial^{j}\partial_j a_{i}-\partial_{i}\partial^{j}a_{j}]=0,~[ M^{(n)}_{k},\partial^{i}\partial_{i} a_{t}-\partial_{t}\partial^{i}a_{i}]=0.
	\eea
	From \eqref{infchecka} and \eqref{infcheckma}, we see that Galilean electrodynamics is invariant under GCA in $d=4$ case. For further explanations, the reader is directed to \cite{Bagchi:2014ysa}.
	
	\subsection{An action for Galilean electrodynamics in $d=4$}\label{sec:null-redux}
	
	In this section, as a warm-up to our analysis in Yang Mills theory, we review the construction of the action for Galilean electrodynamics by null reducing electrodynamics (ED) in one higher dimension \cite{Festuccia:2016caf, Chapman:2020vtn}. For that, let us consider the 5-dimensional Maxwell action 
	\begin{eqnarray}\label{RMA}
	\mathcal{S}_{ED}= \int d^{5}x \bigg(-\frac{1}{4}\eta^{\tilde{\mu}\tilde{\rho}}\eta^{\tilde{\nu}\tilde{\sigma}}F_{\tilde{\mu}\tilde{\nu}}F_{\tilde{\rho}\tilde{\sigma}}\bigg),
	\end{eqnarray}
	in the coordinate system \eqref{ofc2}. The field strength is given by $F_{\tilde{\mu}\tilde{\nu}}=(\partial_{\tilde{\mu}}A_{\tilde{\nu}}-\partial_{\tilde{\nu}}A_{\tilde{\mu}})$. We take the vector potential $A_{\tilde{\mu}}$ to be independent of the null coordinate `$u$' and decompose the components of $A_{\tilde{\mu}}$ as
	\begin{equation}
	A_{u}=\phi, \  A_{t}=a_{t}, \ A_{i}=a_{i}. \label{redux-ansatz-fields}
	\end{equation}
	Then performing null reduction, we get an action for Galilean electrodynamics in $4$-dimensions as
	\begin{eqnarray}\label{GEDL}
	\mathcal{S}_{GED}=\int dt\,d^{3}x\bigg(-\frac{1}{4}W^{ij}W_{ij}+E^i\partial_{i}\phi+\frac{1}{2}(\partial_{t}\phi)^2\bigg),
	\end{eqnarray}
	where $W_{ij}=(\partial_{i}a_{j}-\partial_{j}a_{i})$ and $E_i=(\partial_{t}a_{i}-\partial_{i}a_{t})$. The associated EOM are given by
	\bea{eomaa}
	 \partial^2_{t}\phi+\partial_{i}E_{i}=0,\label{aeom1}\quad \partial^{i}\partial_{i}\phi=0,\label{aeom2}\quad
	 \partial^{j}W_{ji}-\partial_{t}\partial_{i}\phi=0.\label{aeom3}
	\eea
	We see that the action \eqref{GEDL} is invariant under GCA \eqref{gca rep} in $d=4$.
	We further see that setting $\phi=0$, equations \eqref{eomaa} reduce to 
		\begin{eqnarray}\label{eomaaphiiszeroabelian}
		&& \partial^{j}W_{ji}=0, \quad \partial_{i}E_{i}=0.
	\end{eqnarray}
which are the EOM in the magnetic limit of GED \eqref{Meom} \cite{Bagchi:2014ysa}. Thus we see that from a null reduction, we only get the magnetic sector of GED.

	\subsection{Galilean electrodynamics in $d=3$ dimensions}
	We will study Galilean electrodynamics in $3$-dimensions obtained by null reduction of electrodynamics in $4$-dimensions. We see that the action looks identical to the case in four dimensions \eqref{GEDL} with the difference being in the values of spatial indices $i,j$. However the symmetries are different: GED in $3$-dimensions is invariant under Schr\"{o}dinger algebra \cite{Chapman:2020vtn}.

\medskip
	
We will now look at the invariance of the GED action under the Schr\"{o}dinger symmetries. Under boost transformations, using the transformations of the fields \eqref{bootsa}, the action transforms as
\begin{equation}
\delta_{G_{k}} \mathcal{S}_{GED} =\int dt\,d^2x\, \delta_{G_{k}}\mathcal{L}_{GED} = \int dtd^2x\,\p_k \Big[- t\Big(\frac{1}{2}\p_{t}\phi\p_{t}\phi+E^{i}\p_{i}\phi-\frac{1}{4}W^{ij}W_{ij}\Big)\Big],
\end{equation}
	where $\mathcal{L}_{GED}= \Big(\frac{1}{2}\p_{t}\phi\p_{t}\phi+E^{i}\p_{i}\phi-\frac{1}{4}W^{ij}W_{ij}\Big)$ is the Lagrangian density.	 Under dilatation, the fields transform as given in \eqref{stsa} with scaling dimensions $(\Delta_{\phi}, \Delta_{a_t}, \Delta_{a_t})$, to be determined. We see that using these transformations of the fields, the action transforms to a total derivative
\begin{equation}
	\delta_{\tilde{D}} \mathcal{S}_{GED}=\int dt\,d^2 x\,\delta_{\tilde{D}}\mathcal{L}_{GED}
	= \int d^2x\Big[\partial_{t}(2 \,t\mathcal{L}_{GED})+\partial_{k}(x^{k}\mathcal{L}_{GED})\Big]
\end{equation}
	if the scaling dimensions of the fields are taken to be
	\begin{equation}
		\Delta_{\phi}=0,\,\Delta_{a_{i}}=1,\,\Delta_{a_t}=2. \label{3d-GED-scaling-dim}
	\end{equation}
Under special conformal transformations, using \eqref{sctsa} the action changes as
\begin{eqnarray}
\delta_{\tilde{K}} \mathcal{S}_{GED}=\int dtd^2x\Big[\p_{t}(2t^{2}\mathcal{L}_{GED})+\p_{k}(2tx^{k}\mathcal{L}_{GED})\Big].
\end{eqnarray}
We see that the action transforms to a total derivatives under boosts, dilatation and the the special conformal transformation. Thus, the action is invariant under the Schr\"{o}dinger symmetry in $d=3$ dimensions.

\medskip

\noindent Now we will show that the GED action is invariant under the infinite symmetries generated by $Y_i^{(n)}$ and the global symmetries generated by $Z^{(n)}$ for $n=-1,0,1$. To see this, we use the transformations of the fields \eqref{3d L transformation abelian1}, \eqref{3d M trans abelian1} under $Y_{i}^{(n)}$ and $Z^{(n)}$, and get the changes in the action as
\begin{eqnarray}
	&& \delta_{Z^{(n)}} \mathcal{S}_{GED}=\int dtd^2x \Big[\partial_{t}\Big(2t^{n+1}\mathcal{L}_{GED}\Big)+\partial_{k}\Big(\,(n+1)t^{n}x^{k}\mathcal{L}_{GED}\Big) \non \\
	&& \hspace{40mm} -n(n+1)(n-1)\,t^{n-2}(\phi)^2\Big], \nonumber \\
	&& \delta_{Y^{(n)}_{k}} \mathcal{S}_{GED} =\int dt\,d^2x\,\partial_{k}\Big[- \Big(t^{n+\frac{1}{2}}\,\mathcal{L}_{GED}-\frac{1}{2}(n-\frac{1}{2})(n+\frac{1}{2})t^{n-\frac{3}{2}}(\phi)^2\Big)\Big].
\end{eqnarray}
We see that $\delta_{Y_{k}^{(n)}}\mathcal{S}_{GED}$ is a total derivative for all $n$, and $\delta_{Z^{(n)}}\mathcal{S}_{GED}$ is a total derivative only for $n=(-1,0,1)$, implying invariance under $Z^{(-1,0,1)}\equiv(H,D,K)$.

The next thing we are interested in is looking at the symmetries of the EOM \eqref{eomaa} of the theory. The analysis here is very similar to what we will present later on in the Yang-Mills case. Instead of repeating ourselves, we point the reader to Sec~\ref{gym3d} and all the results would hold by putting the structure constants to zero. The upshot is that the EOM are invariant under the finite Schr{\"o}dinger algebra as well as the infinite dimensional boosts, as in the case of the action. However, like in the $d=4$ GED case, when the field $\phi$ is turned off, the EOM exhibit an extended symmetry and remain invariant under the full infinite Schr{\"o}dinger-Virasoro algebra.

	\subsection{Electromagnetic duality}
	
	Earlier in this section, we encountered the electric-magnetic duality in 4$d$ GED which exchanges the electric and magnetic sectors \eqref{4d-GED-em-duality}. Then it is natural to ask if there is some notion of similar duality in 3$d$ GED obtained by null reduction. Given the result in 4$d$ GED, this is not obvious since we get only the magnetic sector from null reduction. As we will see in this section, there are electromagnetic type dualities in GED in $3$-dimensions, and these are of different kind than the electric-magnetic duality in $4$-dimensions. We will heavily use null reduction in the following.
	
	\subsubsection{Free theory}
	The electromagnetic duality in relativistic electrodynamics in $4$-dimensions is the invariance of Maxwell's equations $\partial_{\mu}F^{\mu\nu}=0$ and Bianchi identity $\p_{\mu}\tilde{F}^{\mu\nu}=0$ under the duality transformation
	\begin{equation}
	F^{\mu\nu} \rightarrow F'^{\mu\nu}=\tilde{F}^{\mu\nu}, \quad \tilde{F}^{\mu\nu} \rightarrow \tilde{F}'^{\mu\nu}=F^{\mu\nu}. \label{em-duality-4d}
	\end{equation}
	Here the dual of the field strength is defined, in our conventions, as
	\begin{equation}\label{Dual Relativistic}
	\tilde{F}^{\mu\nu}=\frac{1}{2}\epsilon^{\mu\nu\rho\sigma}F_{\rho\sigma},
	\end{equation}
	where $\epsilon^{\mu\nu\rho\sigma}$ is the totally antisymmetric tensor with $\epsilon^{0123} = 1$.
	
	\medskip
	
	\noindent To get electromagnetic duality in GED in $3$-dimensions, let us perform null reduction using the reduction ansatz \eqref{ofc2}, \eqref{redux-ansatz-fields}. The field strength reduces to field strength like variables in the Galilean theory as
	\begin{eqnarray}
	&& W_{t} = F_{ut} =-\partial_{t}\phi, \quad W_{i} = F_{ui} =-\partial_{i}\phi, \quad W_{it}=F_{it}=\p_{i}a_{t}-\p_{t}a_{i}, \nonumber \\
	&& W_{ij}=F_{ij}=\p_{i}a_{j}-\p_{j}a_{i}, \label{Galilean-field-strength}
	\end{eqnarray}
	and the dual field strength reduces to dual field strength like variables in the Galilean theory as
	\begin{eqnarray}
	&& \tilde{W}^{t}=\tilde{F}^{ut}=\frac{1}{2}\e^{tij}W_{ij}, \quad \tilde{W}^{i}=\tilde{F}^{ui}=-\e^{tij}W_{tj}, \quad \tilde{W}^{it}=\tilde{F}^{it}=-\e^{tij}W_{j}, \nonumber \\
	&& \tilde{W}^{ij}=\tilde{F}^{ij}=\e^{tij}W_{t}. \label{Galilean-dual-field-strength}
	\end{eqnarray}
	Here $\epsilon^{tij}$ is totally antisymmetric with $\epsilon^{t23} = -1$ and it is related to the totally antisymmetric tensor in $4$-dimensions as $\epsilon^{utij} = \epsilon^{tij}$. Then the Galilean EOM can be written in terms of the field strength like variables as
	\begin{eqnarray}\label{3d-GED-eom}
	\delta^{ij}\p_{i}W_{j}=0, \quad \delta^{jk}\p_{j}W_{ki}+\p_{t}W_{i}=0, \quad \p_{t}W_{t}+\delta^{ij}\p_{i}W_{jt}=0.
	\end{eqnarray}
	The null reduction of the Bianchi identity $\partial_{\mu}\tilde{F}^{\mu\nu} = 0$ gives Bianchi identities in the Galilean theory as
	\begin{eqnarray}\label{3d-GED-bianchi}
	\p_{t}\tilde{W}^{t}+\p_{i}\tilde{W}^{i}=0, \quad \p_{i}\tilde{W}^{it}=0, \quad \p_{t}\tilde{W}^{ti}+\p_{k}\tilde{W}^{ki}=0.
	\end{eqnarray}
	Now to do null reduction of the relativistic electromagnetic duality transformation \eqref{em-duality-4d}, let us first express it in terms of field strength $F_{\mu\nu}$ and dual field strength $\tilde{F}^{\mu\nu}$ with proper index structure:
	\begin{eqnarray}\label{dtnr1}
	F'_{\alpha\beta}=\eta_{\alpha\mu}\eta_{\beta\nu}\tilde{F}^{\mu\nu}, \quad \tilde{F}'^{\mu\nu}=\eta^{\mu\alpha}\eta^{\beta\nu}F_{\alpha\beta}.
	\end{eqnarray}
	Null reducing these expressions, we get the duality transformations in Galilean theory, $W \rightarrow W'=\tilde{W} $, $\tilde{W}\rightarrow \tilde{W}'=W$ as
	\begin{eqnarray}
	&& W_{t}\rightarrow W'_{t}=-\tilde{W}^{t}, \quad W_{i}\rightarrow W'_{i}=\delta_{ij}\tilde{W}^{tj}, \quad W_{ti}\rightarrow W'_{ti}=\delta_{ij}\tilde{W}^{j}, \nonumber \\
	&& W_{ij}\rightarrow W'_{ij}=\delta_{ik}\delta_{jl}\tilde{W}^{kl}, \nonumber \\
	&& \tilde{W}^{t}\rightarrow \tilde{W}'^{t}=-W_{t}, \quad \tilde{W}^{i}\rightarrow \tilde{W}'^{i}=\delta^{ij}W_{tj}, \quad \tilde{W}^{ti}\rightarrow \tilde{W}'^{ti}=\delta^{ij}W_{j}, \nonumber \\
	&& \tilde{W}^{ij} \rightarrow \tilde{W}'^{ij}=\delta^{ik}\delta^{jl}W_{kl}. \label{em-duality-3d-GED}
	\end{eqnarray}
	We see that under the transformations \eqref{em-duality-3d-GED}, the Galilean EOM \eqref{3d-GED-eom} for $(W_t, W_i, W_{ti}, W_{ij})$ become Galilean Bianchi identities \eqref{3d-GED-bianchi} for $(\tilde{W}'^t, \tilde{W}'^i, \tilde{W}'^{ti}, \tilde{W}'^{ij})$ and the Galilean Bianchi identities for $(\tilde{W}^t, \tilde{W}^i, \tilde{W}^{ti}, \tilde{W}^{ij})$ become Galilean EOM for $(W'_t, W'_i, W'_{ti}, W'_{ij})$. Thus the field equations \eqref{3d-GED-eom} and \eqref{3d-GED-bianchi} are invariant under the transformations \eqref{em-duality-3d-GED}.

	\subsubsection{Interacting theory}
	
	We have a generalization of the electromagnetic duality to $SL(2,\mathbb{R})$ duality in a relativistic interacting theory of a complex scalar $Z(x)$ and the electromagnetic field $A_{\mu}(x)$ in $4$-dimensions \cite{Freedman:2012zz}. This theory is described by the Lagrangian density
	\begin{equation}
	\mathcal{L}_{ED}=-\frac{1}{4}(ImZ)F_{\mu\nu}F^{\mu\nu}-\frac{1}{8}(ReZ)\e^{\mu\nu\rho\sigma}F_{\mu\nu}F_{\rho\sigma}, \label{rel-scalarED-lagrangian}
	\end{equation}
	and the EOM
	\begin{equation}\label{scalarED-ME}
	\p_{\mu}\big[(ImZ)F^{\mu\nu}+(ReZ)\tilde{F}^{\mu\nu}\big]=0,
	\end{equation}
	along with the Bianchi identity. These EOM and Bianchi identity are invariant under $SL(2,\mathbb{R})$ duality transformations
	\begin{eqnarray}\label{SL2R-transformation}
	\mathcal{S}=\bigg(\begin{matrix}
	d & c\\
	b & a
	\end{matrix}\bigg), \quad \quad ad-bc=1,
	\end{eqnarray}
	with the fields transforming as
	\begin{eqnarray}&&
	F_{\mu\nu}\rightarrow F'_{\mu\nu}=\big[d+\frac{c}{2}(Z+\bar{Z})\big]F_{\mu\nu}+\frac{ic}{2}(Z - \bar{Z})\tilde{F}^{\alpha\beta}\eta_{\alpha\mu}\eta_{\beta\nu}, \nonumber \\&&
	\tilde{F}^{\mu\nu}\rightarrow \tilde{F}'^{\mu\nu}=\big[d+\frac{c}{2}(Z+\bar{Z})\big]\tilde{F}^{\mu\nu} -\frac{ic}{2}(Z-\bar{Z}){F}_{\alpha\beta}\eta^{\alpha\mu}\eta^{\beta\nu}, \label{SL2R-transformations}
	\end{eqnarray}
	and
	\begin{equation}\label{Z-transformation}
	Z' = \frac{aZ+b}{cZ+d}.
	\end{equation}
	We note that this duality is a symmetry of the EOM and the Bianchi identity and not of the Lagrangian \eqref{rel-scalarED-lagrangian}, as can be checked by a straightforward calculation \cite{Freedman:2012zz}.

	\medskip

	\noindent Performing null reduction of \eqref{rel-scalarED-lagrangian}, we get the Lagrangian describing a theory of complex scalar and GED in $3$-dimensions:
	\bea{}
	\mathcal{L_{M+S}} &= & (\text{Im}Z)\big[\frac{1}{4}W^{ij}W_{ij}+W^i W_{ti}+\frac{1}{2}W_{t}W_{t}\big] \nonumber \\
	&& -\frac{(\text{Re} Z)}{2}\big[\tilde{W}^{t}W_{t}+\tilde{W}^{i}W_{i}+\tilde{W}^{it}W_{it}+\frac{1}{2}\tilde{W}^{ij}W_{ij}\big].
	\eea
	The null reduction of \eqref{scalarED-ME} gives the EOM
	\begin{eqnarray}\label{Abelian+Scalar interacting}
	&& \p_{t}\big[i(Z-\bar{Z})W_{t}+(Z+\bar{Z})\tilde{W}^{t}\big]+\p_{i}\big[i(Z-\bar{Z})\delta^{ij}W_{jt}+(Z+\bar{Z})\tilde{W}^{i}\big]=0, \nonumber
	\\
	&& \p_{i}\big[i(Z-\bar{Z})\delta^{ij}W_{j}+(Z+\bar{Z})\tilde{W}^{it}\big]=0, \label{GMNreduction} \\
	&& \p_{t}\big[i(Z-\bar{Z})\delta^{ij}W_{j}-(Z+\bar{Z})\tilde{W}^{ti}\big]+\p_{k}\big[i(Z-\bar{Z})\delta^{kl}\delta^{ij}W_{lj}-(Z+\bar{Z})\tilde{W}^{ki}\big]=0. \nonumber
	\end{eqnarray}
	The Bianchi identities are same as given in \eqref{3d-GED-bianchi}. Now we perform null reduction on the action of $SL(2,\mathbb{R})$ transformations on relativistic fields \eqref{SL2R-transformations}. This null reduction, using \eqref{Galilean-field-strength}, \eqref{Galilean-dual-field-strength}, gives the action of $SL(2,\mathbb{R})$ transformations on the fields in the Galilean theory as
	\begin{eqnarray}&&
	W_t \ \rightarrow\  W'_{t}=\big[d+\frac{c}{2}(Z+\bar{Z})\big]W_{t}-\frac{ic}{2}(Z-\bar{Z})\tilde{W}^{t}, \nonumber \\&&
	W_i \ \rightarrow\  W'_{i}=\big[d+\frac{c}{2}(Z+\bar{Z})\big]W_{i}+\frac{ic}{2}(Z-\bar{Z})\tilde{W}^{tj}\delta_{ji}, \nonumber \\&&
	W_{it} \ \rightarrow\  W'_{it}=\big[d+\frac{c}{2}(Z+\bar{Z})\big]W_{it}-\frac{ic}{2}(Z-\bar{Z})\tilde{W}^{j}\delta_{ji}, \nonumber \\&&
	W_{ij} \ \rightarrow\  W'_{ij}=\big[d+\frac{c}{2}(Z+\bar{Z})\big]W_{ij}+\frac{ic}{2}(Z-\bar{Z})\tilde{W}^{kl}\delta_{ki}\delta_{lj}, \label{SL2R-f-transformations}
	\end{eqnarray}
	and
	\begin{eqnarray}&&
	\tilde{W}^{t}\ \rightarrow\ \tilde{W}'^{t}=\big[d+\frac{c}{2}(Z+\bar{Z})\big]\tilde{W}^{t} +\frac{ic}{2}(Z-\bar{Z})W_{t}, \nonumber \\&&
	\tilde{W}^{i}\ \rightarrow\ \tilde{W}'^{i}=\big[d+\frac{c}{2}(Z+\bar{Z})\big]\tilde{W}^{i} -\frac{ic}{2}(Z-\bar{Z})W_{tj}\delta{ji}, \nonumber \\&&
	\tilde{W}^{it}\ \rightarrow\ \tilde{W}'^{it}=\big[d+\frac{c}{2}(Z+\bar{Z})\big]\tilde{W}^{it} + \frac{ic}{2}(Z-\bar{Z})W_{j}\delta^{ji}, \nonumber \\&&
	\tilde{W}^{ij}\ \rightarrow\ \tilde{W}'^{ij}=\big[d+\frac{c}{2}(Z+\bar{Z})\big]\tilde{W}^{ij} - \frac{ic}{2}(Z-\bar{Z})W_{kl}\delta^{ki}\delta^{lj}. \label{SL2R-tildef-transformations}
	\end{eqnarray}
	To see the invariance of the field equations, we can first replace fields $(W,\tilde{W})$ by $(W',\tilde{W}')$ in \eqref{GMNreduction},\eqref{3d-GED-bianchi}, \emph{i.e.} assume that $(W',\tilde{W}')$ satisfy the field equations and the Bianchi identities. Then using the transformations \eqref{SL2R-f-transformations}, \eqref{SL2R-tildef-transformations} and $ad-bc=1$, we can show that the fields $(W,\tilde{W})$ satisfy equations \eqref{GMNreduction} and \eqref{3d-GED-bianchi}.

\medskip

The takeaway point from this analysis is that electromagnetic duality in Galilean electrodynamics is very different in four and three dimensions. While in four dimensions, there is an exchange between the electric and magnetic sectors, which are in effect two different theories, in three dimensions, the duality exists already in one sector, which reduces to the magnetic sector by putting an extra field to zero. A priori, this may seem like a surprise, as the actions of both the 4d and the 3d theories are almost identical and obtained in the same way from the higher dimensional relativistic theory. The answer to the difference is of course that the original 5d relativistic theory from which the 4d action is derived, is not invariant under electromagnetic duality, whereas the 4d relativistic theory is. The duality structure of the 3d Galilean theory is directly inherited from the 4d relativistic theory through null reductions. 

\bigskip
	
\section{Non-Abelian Galilean Gauge theory}
In this section, we move on to the non-Abelian Galilean gauge theories. Before doing so, in order to set notation, let us very briefly recall the relativistic theory and the symmetries associated with it.

It is well known that classical relativistic Yang-Mills theory is invariant under conformal symmetry in four spacetime dimensions. The Yang-Mills action in $(d+1)$-dimensions is:
	\begin{eqnarray}\label{yma}
	\mathcal{S}_{YM} = \int d^{d+1}x	\,\mathcal{L}_{YM}= \int d^{d+1}x\,\Big(-\frac{1}{4}F^{\tilde{\mu} \tilde{\nu} a}F_{\tilde{\mu} \tilde{\nu}}^{ a} \Big),
	\end{eqnarray}
	and the EOM
	\begin{eqnarray}\label{eomym}
	\partial_{\tilde{\mu}}F^{\tilde{\mu}\tilde{\nu} a}+gf^{abc}A_{\tilde{\mu}}^{b}F^{\tilde{\mu}\tilde{\nu} c}=0,
	\end{eqnarray}
	where $a=1,2,...,N^2 -1$. The non-abelian field strength is defined as $F_{\tilde{\mu} \tilde{\nu}}^a=\p_{\tilde{\mu}}A_{\tilde{\nu}}^a-\p_{\tilde{\nu}}A_{\tilde{\mu}}^a+gf^{abc}A_{\tilde{\mu}}^bA_{\tilde{\nu}}^c$. Here, $A^{a}_{\tilde{\mu}}$ is the gauge field and $f^{abc}$ is the structure constant of the underlying gauge group. The action \eqref{yma} and equations \eqref{eomym} are trivially invariant under Poincar{\'e} and scale transformations in all dimensions.
\begin{eqnarray}&&
	\delta_{K_{\tilde{\sigma}}}\big(\partial_{\tilde{\mu}}F^{\tilde{\mu}\tilde{\nu} a}+gf^{abc}A_{\tilde{\mu}}^{b}F^{\tilde{\mu}\tilde{\nu} c}\big)=\big(d-3\big)\big[F_{\tilde{\sigma}\tilde{\nu}}^{a}+\big(\p_{\tilde{\sigma}}A^{a}_{\tilde{\nu}}-\eta_{\tilde{\sigma}\tilde{\nu}}\p_{\tilde{\rho}}A^{\tilde{\rho} a}\big)
	+gf^{abc}\bigl\{2A^{b}_{\tilde{\sigma}}A_{\tilde{\nu}}^{c}\non\\&&\hspace{7cm}-\eta_{\tilde{\sigma}\tilde{\nu}}A^{\tilde{\rho} b}A_{\tilde{\rho}}^{c}+\eta_{\tilde{\sigma}\tilde{\rho}}x^{\tilde{\rho}}\big(\p^{\tilde{\rho}}(A^{b}_{\tilde{\rho}}A^{c}_{\tilde{\nu}})\non\\&&\hspace{7cm}+A^{\tilde{\rho} b}F_{\tilde{\rho}\tilde{\nu}}^{c}+gf^{cde}A^{\tilde{\rho} b}A^{d}_{\tilde{\rho}}A^{e}_{\tilde{\nu}}\big)\bigr\}\big].
	\end{eqnarray}
We see that the EOM are invariant under SCT in $d+1=4$ spacetime dimensions. So relativistic Yang-Mills theory is invariant under the entire conformal group at the classical level in four spacetime dimensions. 
	
	\subsection{Galilean Yang-Mills Theory}
We begin our investigation of non-Abelian gauge theories with a brief review of existing literature. In \cite{Bagchi:2015qcw}, a subset of us  looked at the generalised $SU(N)$ Galilean field theory.  Below we review the simplest case, i.e. the $SU(2)$ theory. The first non-trivial generalisation is the existence of skewed limits when one deviates from a $U(1)$ theory. It is due to the presence of three different gauge fields that leads to four limits instead of two in the $U(1)$ case. The scaling of gauge fields ($A_{\mu}^a=a_{t}^a, a_{i}^a$) are defined as
	\bes{}
	\bea{sca} \mbox{{\bf{Electric limit}}:}\,&& a^{a}_t \rightarrow a^{a}_t, \, \, a^{a}_i \rightarrow \epsilon a^{a}_i ,\\\mbox{{\bf{Magnetic limit}}:}\,&&
	a^{a}_t \rightarrow \epsilon a^{a}_t, \, \, a^{a}_i \rightarrow a^{a}_i,\\
	\mbox{{\bf{EEM limit}}:}\,&& a^{1,2}_t \rightarrow a^{1,2}_t, \, \, a^{1,2}_i \rightarrow \epsilon a^{1,2}_i,~ a^{3}_t \rightarrow \epsilon a^{3}_t, \, \, a^{3}_i \rightarrow a^{3}_i,\\
	\mbox{{\bf{EMM limit}}:}\,&& a^{1}_t \rightarrow a^{1}_t, \, \, a^{1}_i \rightarrow \epsilon a^{1}_i,~ a^{2,3}_t \rightarrow \epsilon a^{2,3}_t, \, \, a^{2,3}_i \rightarrow a^{2,3}_i.\eea \ees
	Here, $a=(1,2,3)$ and $E$ denotes for electric and $M$ for magnetic case. When we apply these four sets of scaling on the Non-abelian EOM \eqref{eomym}, we will get the vanilla limits (Electric and Magnetic limits) as well as the skewed limits (EEM and EMM limits). We will now write down the EOM for each sector. They are given by
	\begin{itemize}
		\item \underline{Electric limit}:
		\begin{eqnarray}
		\p_{i}\p_{i}a_{i}^{a}-\p_{j}\p_{i}a_{j}^{a}+\p_{t}\p_{i}a_{t}^{a}+g\e^{abc}a_{t}^{b}\p_{i}a_{t}^{c}=0,\quad \p_{i}\p_{i}A_{t}^{a}=0.
		\end{eqnarray}
		\item \underline{Magnetic limit}:\be{MMMeom}
		\partial^{i}\partial_{i}a^{a}_{t}-\partial^{i}
		\partial_{t}a^{a}_{i}=0, \quad
		\partial^{j}\partial_{j} a^{a}_{i}-\partial^{j}\partial_{i}a^{a}_{j}=0.
		\ee
		\item{}\underline{EEM limit}:
		\bes \label{EEMeom}
		\bea{}
		&& \partial^{i}\partial_{i} a_{t}^{1,2}=0,\quad \quad
		\partial^{i}\partial_{i}a_{j}^{1,2}
		-\partial^{i}\partial_{j}a_{i}^{1,2}
		+\partial_{t}\partial_{j}a_{t}^{1,2}=0,\\
		&& \partial^{i}\partial_{i} a_{t}^{3}-\partial^{i}\partial_{t}a_{i}^{3}=0,\quad
		\partial^{i}(\partial_{i}a_{j}^{3}-\partial_{j}a_{i}^{3})=0.
		\eea
		\ees
		\item \underline{EMM limit}:
		\bes\label{EMMeom}
		\bea{}
		\partial^{i}(\partial_{i}a_{j}^{1}-\partial_{j}a_{i}^{1})
		+\partial_{t}\partial_{j}a_{t}^{1} +g \partial^{i}(a^{2}_{i}a^{3}_{j}-a^{3}_{i}a^{2}_{j}) && \non\\
		+ g a^{2}_{i}(\partial_{i}a^{3}_{j}-\partial_{j}a^{3}_{i})+ g a^{3}_{i}(\partial_{j}a^{2}_{i}-\partial_{i}a^{2}_{j}) &=& 0, \label{eom1}
		\\
		\non\\
		\partial^{i}\partial_{i} a^{1}_{t}=0, \quad
		\partial^{i}(\partial_{i}a_{j}^{2,3}-\partial_{j}a_{i}^{2,3})&=&0, \label{eom2}\\
		\non\\
		\partial^{i}\partial_{i} a_{t}^{3}-\partial^{i}\partial_{t}a_{i}^{3} -2g a_{i}^{2}\partial^{i}a_{t}^{1}
		-gA_{t}^{1}\partial^{i}a_{i}^{2}&=&0, \label{eom3}\\
		\partial^{i}\partial_{i} a_{t}^{2}-\partial^{i}\partial_{t}a_{i}^{2} +2g a_{i}^{3}\partial^{i}a_{t}^{1}+ga_{t}^{1}
		\partial^{i}a_{i}^{3}&=&0. \label{eom4}
		\eea
		\ees
	\end{itemize}
	For $SU(2)$ theory, the structure constant $f^{abc}\equiv \varepsilon^{abc}$ with $\varepsilon^{abc}= \pm 1$ for different permutations. In \cite{Bagchi:2015qcw}, the invariance of EOM of these four limits of Galilean Yang-Mills was seen under infinite-dimensional GCA.
	
Below we will obtain a different set of EOM from the ones above by looking at a null reduction of relativistic Yang-Mills in $d=4$. Needless to say, turning the coupling $g=0$, and setting an extra field to zero, like in the electrodynamics case, we will recover the Magnetic sector EOM above \refb{eom2}. The magnetic sector in the above is the same as the $U(1)$ case, with some added gauge indices. So given the abelian null-reduced Galilean theory reproduces the Magnetic equations, it is a foregone conclusion that turning off the gauge coupling (and setting the extra field to zero) would reduce the EOM to \refb{eom2}. What is unexpected is that the EOM from the action we derive does not fit into any one we derived using the limiting procedure. The infinite dimensional symmetries, however, would still emerge from this new set of equations.

	\subsection{Null Reduction of Yang-Mills theory}
	We will construct the action for Galilean Yang-Mills theory by using the null reduction procedure. We will follow the same technique that was applied to scalar field theory and Maxwell theory to get their Galilean counterparts. We write the Lagrangian density of Yang-Mills theory in $(d+1)$ dimensions \eqref{yma} in null coordinates \eqref{ofc2}:
	\begin{eqnarray}\label{hjed}
	\mathcal{L}_{YM}= -\frac{1}{4}\eta^{\tilde{\mu}\tilde{\rho}}\eta^{\tilde{\nu}\tilde{\sigma}}F_{\tilde{\mu}\tilde{\nu}}^{a}F_{\tilde{\rho} \tilde{\sigma}}^{a}=-\frac{1}{4}\Big[2F^{a}_{ut}F^{a}_{tu}+F^{ija}F^{a}_{ij}+4F^{a}_{ui}F^{ia}_{t}\Big],
	\end{eqnarray}
	and perform null reduction along the null direction parametrized by the coordinate $u$. We take the gauge field to be independent of the $u$-coordinate, \ie\ $\partial_u A_{\tilde{\mu}}^a = 0$, and decompose its components as
	\begin{eqnarray}\label{frn}
	A_{u}^{a}=\phi^a,\quad A_{t}^a=a_{t}^a,\quad A_{i}^{a}=a_{i}^{a}.
	\end{eqnarray}
	Then the null reduction gives the Lagrangian density in $d$ spacetime dimensions as
	\bea{}&&\hspace{-1.5cm}
	\mathcal{L}_{GYM}=\Big[\frac{1}{2}(\partial_{t}\phi^{a}-gf^{abc}\phi^{b}a_{t}^{c})(\partial_{t}\phi^{a}-gf^{ade}\phi^{d}a_{t}^{e})-\frac{1}{4}(\partial^{i}a^{j}-\partial^{j}a^{i}+gf^{ade}a^{id}a^{je})\non\\&&(\partial_{i}a_{j}-\partial_{j}a_{i}+gf^{abc}a_{i}^{b}a_{j}^{c}) 
	+(\partial_{i}\phi^{a}-gf^{abc}\phi^{b}a_{i}^{c})(\partial_{t}a^{ia}-\partial^{i}a_{t}^{a}+gf^{abc}a_{t}^{b}{a}^{ic})\Big],
	\eea
	where $GYM$ stands for Galilean Yang-Mills. It can also be written in a compact form given by 
	\begin{eqnarray}\label{ngym}
	\mathcal{L}_{GYM}=\frac{1}{2}D_{t}\phi^{a}D_{t}\phi^{a}+D_{i}\phi^{a}E^{ia}-\frac{1}{4}W^{ija}W_{ij}^{a},
	\end{eqnarray}
	where $D_t$, $D_i$ are gauge-covariant derivatives and $E^{ia}$, $W^a_{ij}$ are field strength variables defined as
	\bes{}\label{quan}
	\begin{eqnarray}\label{Covariant form}&&
	D_{t}\phi^{a}=\partial_{t}\phi^{a}-gf^{abc}\phi^{b}a_{t}^{c},~
	D_{i}\phi^{a}=\partial_{i}\phi^{a}-gf^{abc}\phi^{b}a_{i}^{c},\\&&
	E^{ia}=\partial_{t}a^{ia}-\partial^{i}a_{t}^{a}+gf^{abc}a_{t}^{b}{a}^{ic},~
	W_{ij}^{a}=\partial_{i}a_{j}-\partial_{j}a_{i}+gf^{abc}a_{i}^{b}a_{j}^{c}.
	\end{eqnarray}
	\ees
	The EOM for the Lagrangian \eqref{ngym} are given by
	\bes{}\label{EOM of Yang-Mills}
	\begin{eqnarray}&&
	D_{t}D_{t}\phi^{a}+D_{i}E^{ia}=0,\label{EOM-1}\\&&
	D_{i}D_{i}\phi^{a}+gf^{abc}\phi^{b}D_{t}\phi^{c}=0,\label{EOM2}\\&&
	D_{t}D_{i}\phi^{a}-D_{j}W_{ji}^{a}-gf^{abc}\phi^{b}E^{ic}=0.\label{EOM3}
	\end{eqnarray}\ees
	We can also find these equations by doing the procedure of null reduction on relativistic equations.\footnote{The Lagrangian was also introduced in \cite{Gomis:2020fui} to derive the EOM for Galilean Yang-Mills theory with $U(N)$ gauge group obtained as an effective theory from non-relativistic open string theory.}
	
	\medskip
	
	\noindent If we put $\phi^{a}=0$, the equations become
	\begin{eqnarray}\label{phi0}
	D_{i}E^{ia}=0,~~
	D_{j}W_{ji}^{a}=0.
	\end{eqnarray}

	\subsection{Symmetries of Galilean Yang-Mills theory in $d=4$}\label{sec:GYM-symmetries}
	We will explicitly look at the symmetries of the Galilean Yang-Mills at the level of Lagrangian and EOM in $d=4$. We will use the action of GCA to find the symmetries of \eqref{ngym} and \eqref{EOM of Yang-Mills}. In the representation theory, we have some undefined constants given as $(\Delta,a,r,s)$. They depend on the fields of the theory under consideration. The scaling weight ($\Delta$) for a particular field gets fixed when we impose invariance of the Lagrangian under scale transformation. Similarly, the constants $(a,r,s)$ get fixed by comparing the results from \eqref{gcamn} with $n=0$ and taking the non-relativistic limit on boost transformations for a particular field. For our case, the values of these constants are given by
	\bea{}\label{valco}\Delta_{\phi}=\Delta_{a_i}=\Delta_{a_t}=1,~a=0,r=-1,s=1.\eea
	In the coming section, we will use these values of the constants to find the invariance of the Lagrangian and EOM under GCA.

	\subsubsection{Gauge invariance of Action}
	Before looking at spacetime symmetries of the Lagrangian, we will first discuss in details the gauge transformations. We will evaluate non-relativistic gauge transformation by doing a null reduction on relativistic transformations. We will begin with relativistic gauge transformations given by
	\bea{} A_{\mu}^a\rightarrow A_{\mu}^a+\frac{1}{g}\p_{\mu}\a^a+f^{abc}A^{b}_{\mu}\a^c.\eea
	Performing a null reduction along $u$ direction and using \eqref{frn}, the final result comes out to be
	\bes{}\label{gtf}
	\begin{eqnarray}\label{Galilean Gauge Transformation}&&
	a_{i}^{a}\rightarrow a_{i}^{a}+\frac{1}{g}\partial_{i}\alpha^{a}+f^{abc}a_{i}^{b}\alpha^{c},\\&&
	a_{t}^{a}\rightarrow a_{t}^{a}+\frac{1}{g}\partial_{t}\alpha^{a}+f^{abc}a_{t}^{b}\alpha^{c},\\&&
	\phi^{a}\rightarrow \phi^{a}+f^{abc}\phi^{b}\alpha^{c}.
	\end{eqnarray}\ees
	Here, the field $\phi^{a}$ now transforms as a scalar field in adjoint representation. Let us see how the Lagrangian changes under these transformations. The quantities \eqref{quan} transform as
	\bes{}\label{cquan}
	\begin{eqnarray}&&
	D_{t}\phi^{a} \rightarrow D_{t}\phi^{a}+f^{abc}\alpha^{c}D_{t}\phi^{b},~~
	D_{i}\phi^{a} \rightarrow D_{i}\phi^{a}+f^{abc}\alpha^{c}D_{i}\phi^{b},\\&&
	E^{ia} \rightarrow E^{ia}+f^{abc}\alpha^{c}E^{ib},~~
	W_{ij}^{a} \rightarrow W_{ij}^{a}+f^{abc}\alpha^{c}W_{ij}^{b}.
	\end{eqnarray}\ees
	Using \eqref{cquan}, the Lagrangian density change as $\delta \mathcal{L}_{GYM}=0$. It tell us that the action is invariant under the gauge transformations \eqref{gtf}.
	
	\subsubsection{Spacetime symmetries at the level of Action}
	We will first look into the symmetries of the action and then at the level of equations. The action and EOM are trivially invariant under translations $(H,P_i)$ and rotations $(J_{ij})$. We will only show the invariance under boost $(B_i)$, scale transformation $(D)$ and SCT $(K, K_i)$ and then will move on to exhibit the invariance under infinite extension $(L^{(n)}, M^{(n)}_{i})$ of GCA.

\smallskip

 \noindent\textbf{\underline{Boost transformations}:}
		For our case, fields transform under  boosts as 
		\begin{eqnarray}\label{boots}&&
		\delta_{B_{k}} \phi = -(t\partial_{k}\phi),~\delta_{B_{k}} a_{t}^{a}= -(t\partial_{k}a_{t}^{a}+a_{k}^a),~\delta_{B_{k}} a_{i}^{a}=-(t\partial_{k}a_{i}^{a}-\delta_{ik}\phi^{a}),
		\end{eqnarray}
		We can find these transformations by looking at \eqref{gcamn} with $n=0$ and using the values of the constants as given in \eqref{valco}. To compare these transformation to the ones we have in \eqref{gcamn} with $n=0$, we have used the relation $\delta_{\epsilon}\varphi=[\e Q,\varphi(t,x)]$, where
		$Q$ is the generator of the infinitesimal symmetry transformations acting on a generic field $\varphi$ and $\e$ is the symmetry parameter. The action \eqref{ngym} transforms as
		\begin{eqnarray}
		&&\hspace{-1cm}\delta_{B_{k}} S_{GYM} =\int dt\, d^3x\, \Big(\delta_{B_{k}}\mathcal{L}_{GYM}\Big)\non\\&&\hspace{.6cm}= \int dt\,d^3x\,\p_k \Big[- t\Big(\frac{1}{2}D_{t}\phi^{a}D_{t}\phi^{a}+E^{ia}D_{i}\phi^{a}-\frac{1}{4}W^{ija}W_{ij}^{a}\Big)\Big]
		\end{eqnarray}
		We see that the action comes out to be invariant under boost transformation \eqref{boots}. \\\\
\underline{\textbf{Scale transformation}}: We will see how dilatation $(D)$ affects the Lagrangian.
		The field transformations under dilatation is given by
		\begin{eqnarray}
		\delta_{D} \Phi^{a}=\,(t\partial_{t}+x^{k}\partial_{k}+1)\Phi^a ,
		\end{eqnarray}
		where $\Phi^a\equiv(\phi^a, a_i^a, a_t^a)$. The action changes as
		\begin{eqnarray}&&
		\hspace{-1cm}	\delta_{D} S_{GYM}
		= \int dtd^3x\Big[\partial_{t}(\,t\mathcal{L}_{GYM})+\partial_{k}(\, x^{k}\mathcal{L}_{GYM})\Big].
		\end{eqnarray}
		The action is invariant under dilatation.
 \\\\
 \textbf{\underline{$K_i$ transformations}:} 
		The fields $\Phi^a=(\phi^a, a_i^a, a_t^a)$ transforms under $K_i$ as
		\begin{eqnarray}\label{ssct}&&\hspace{-.9cm}
		\delta_{K_{i}} \phi^{a} =-( t^{2}\partial_{i}\phi^{a}),~
		\delta_{K_{i}} a_{t}^{a}=-(t^{2}\partial_{i}a_{t}^{a}+2ta_{i}^{a}),~
		\delta_{K_{i}} a_{j}^{a}=-(t^{2}\partial_{i}a_{j}^{a}-2t\delta_{ij}\phi^{a}).
		\end{eqnarray}
		Using these transformations of the fields, we get the change in the action
		\begin{eqnarray}
		&&\hspace{-1cm}\delta_{K_{i}} S_{GYM} =\int dt\, d^3x\, \Big(\delta_{K_{i}}\mathcal{L}_{GYM}\Big)\non\\&&\hspace{.6cm}= \int dt\,d^3x\,\p_l \Big[- t^2 \Big(\frac{1}{2}D_{t}\phi^{a}D_{t}\phi^{a}+E^{ia}D_{i}\phi^{a}-\frac{1}{4}W^{ija}W_{ij}^{a}\Big)+(\phi^a)^2\Big]\non\\&&\hspace{.6cm}
		=\int dt\,d^3x\,\partial_{l}\Big[- t^2\,\mathcal{L}_{GYM}+(\phi^a)^2\Big].
		\end{eqnarray}
		Again, the action is invariant under $K_i$ transformation.\\\\
 \textbf{\underline{$K$ transformation}:} 
		Under $K$, the fields transforms as 
		\bes{}
		\begin{eqnarray}&&
		\delta_{K} a_{t}^{a}=(t^{2}\partial_{t}a_{t}^{a}+2tx^{k}\partial_{k}a_{t}^{a}+2ta_{t}^{a}+2x^{k}a_{k}^{a}),\\&&
		\delta_{K} a_{i}^{a}=(t^{2}\partial_{t}a_{i}^{a}+2tx^{k}\partial_{k}a_{i}^{a}+2ta_{i}^{a}-2x_{i}\phi^a),\\&&
		\delta_{K} \phi^{a} = (t^{2}\partial_{t}\phi^{a}+2tx^{k}\partial_{k}\phi^{a}+2t\phi^{a}).
		\end{eqnarray}\ees
		We have used \eqref{valco} in \eqref{gcaln} with $n=1$ to get these variations in fields. Using these transformations in action, we get  
		\begin{eqnarray}&&
		\delta_{K} S_{GYM}=\int dt\,d^3x\, \Big[\partial_{t}\Big\{t^2\Big(\frac{1}{2}D_{t}\phi^{a}D_{t}\phi^{a}+E^{ia}D_{i}\phi^{a}-\frac{1}{4}W^{ija}W_{ij}^{a}\Big)+(\phi^{a})^2\Big\}\non\\&&\hspace{4cm}+\partial_{k}\Big\{ 2tx^k \Big(\frac{1}{2}D_{t}\phi^{a}D_{t}\phi^{a}+E^{ia}D_{i}\phi^{a}-\frac{1}{4}W^{ija}W_{ij}^{a}\Big)\Big\}\Big]\non\\&&\hspace{1.5cm}=\int dt\,d^3x\, \Big[\partial_{t}\Big(t^2\mathcal{L}_{GYM}+(\phi^{a})^2\Big)+\partial_{k}\Big(2tx^{k}\mathcal{L}_{GYM}\Big)\Big].
		\end{eqnarray}
		The action comes out to be invariant under $K$ transformation.\\\\
	We have seen that the action is invariant under finite GCA. We will now move on to the infinite extension of GCA. We will see how the action changes and get the required invariance under these transformations \eqref{gca rep}. \\\\
\textbf{\underline{$M_{k}^{(n)}$ transformations}:}
		The fields under $M^{(n)}_{k}$ transform as 
		\bes{}\label{mnk}
		\begin{eqnarray}\label{Mtrans}&&
		\delta_{M^{(n)}_{k}}a_{t}^{a}=-(t^{n+1}\partial_{k}a_{t}^{a}+(n+1)t^{n}a_{k}^{a}),~~\delta_{M^{(n)}_{k}} \phi^{a}=-(t^{n+1}\partial_{k}\phi^{a}),\\&&
		\delta_{M^{(n)}_{k}}a_{i}^{a}=-(t^{n+1}\partial_{k}a_{i}^{a}-(n+1)t^{n}\delta_{ik}\phi^{a}).
		\end{eqnarray}\ees
		If we take $n=0,\pm 1$ in \eqref{mnk}, we get back the global transformations. The action changes as
		\begin{eqnarray}
		&&\hspace{-1cm}	\delta_{M^{(n)}_{k}} S_{GYM} = \int dt\,d^3x\,\p_k \Big[-\, t^{n+1} \Big(\frac{1}{2}D_{t}\phi^{a}D_{t}\phi^{a}+E^{ia}D_{i}\phi^{a}-\frac{1}{4}W^{ija}W_{ij}^{a}\Big)\non\\&&\hspace{7cm}+\frac{1}{2}n(n+1)t^{n-1}(\phi^a)^2\Big]\non\\&&\hspace{.8cm}
		=\int dt\,d^3x\,\partial_{k}\Big[- \Big(t^{n+1}\,\mathcal{L}_{GYM}-\frac{1}{2}n(n+1)t^{n-1}(\phi^a)^2\Big)\Big].
		\end{eqnarray}
		The action becomes invariant under $M^{(n)}_{k}$ transformations.
		\\\\
 \textbf{\underline{$L^{(n)}$ transformations}:} 
		The fields under $L^{(n)}$ transforms as
		\bes{}\label{L transformation}\begin{eqnarray}&&\hspace{-1.2cm}
		\delta_{L^{(n)}} a_{t}^{a}=\Big(t^{n+1}\partial_{t}a_{t}^{a}+(n+1)t^{n}x^{k}\partial_{k}a_{t}^{a}+(n+1)t^{n}a_{t}^{a}+n(n+1)t^{n-1}x^{k}a_{k}^{a}\Big),\\&&\hspace{-1.2cm}
		\delta_{L^{(n)}} a_{i}^{a}=\Big(t^{n+1}\partial_{t}a_{i}^{a}+(n+1)t^{n}x^{k}\partial_{k}a_{i}^{a}+(n+1)t^{n}a_{i}^{a}-n(n+1)t^{n-1}x_{i}\phi^{a}\Big),\\&&\hspace{-1.2cm}
		\delta_{L^{(n)}} \phi^{a}=\Big(t^{n+1}\partial_{t}\phi^{a}+(n+1)t^{n}x^{k}\partial_{k}\phi^{a}+(n+1)t^{n}\phi^{a}\Big).
		\end{eqnarray}\ees
		Under these transformations the action changes as
		\begin{eqnarray}&&\hspace{-1.5cm}
		\delta_{L^{(n)}} S_{GYM}=\int dtd^3x \Big[\partial_{t}\Big(\Big\{t^{n+1}\mathcal{L}_{GYM}+\frac{1}{2}n(n+1)t^{n-1}(\phi^{a})^2\Big\}\Big)\non\\&&\hspace{1cm}+\partial_{k}\Big(\,(n+1)t^{n}x^{k}\mathcal{L}_{GYM}\Big)-n(n+1)(n-1)\,t^{n-2}(\phi^{a})^2\Big].
		\end{eqnarray}
		The action is not invariant under $L^{(n)}$ transformations for all $n$, but invariant only under the global part, \ie\ $L^{(-1,0,1)}\equiv(H,D,K)$.\\\\
	In conclusion, the action is invariant under $M^{(n)}_{k}$ but not invariant under $L^{(n)}$ for all $n$. It comes out to be invariant under the global part of $L^{(-1,0,1)}\equiv(H,D,K)$.

	\subsubsection{Spacetime symmetries at the level of EOM}
	We will now see the invariance at the level of EOM. Here, we will only show the changes of equations under $L^{(n)},M^{(n)}_{k}$. We can find the global part by taking the value of $n=0\pm 1$. \\\\
\textbf{\underline{$M_{k}^{(n)}$ transformations}:} 
		We will use \eqref{mnk} to find the invariance of equations \eqref{EOM of Yang-Mills} under $M_{k}^{(n)}$. We get the result as
		\bes{}
		\begin{eqnarray}&&
		\delta_{M^{(n)}_{k}} \eqref{EOM-1} = -t^{n+1}\partial_{k}\eqref{EOM-1}-(n+1)t^{n}\eqref{EOM3}=0,\\&&
		\delta_{M^{(n)}_{k}} \eqref{EOM3} = -t^{n+1}\partial_{k}\eqref{EOM3}-(n+1)t^{n}\delta_{ik}\eqref{EOM2}=0,\\&&
		\delta_{M^{(n)}_{k}} \eqref{EOM2} = -t^{n+1}\partial_{k}\eqref{EOM2}=0.
		\end{eqnarray}\ees
		Here, \ref{EOM-1}, \ref{EOM2}, etc means that we are considering only the LHS of \eqref{EOM of Yang-Mills}.
		\\\\
\textbf{\underline{$L^{(n)}$ transformations}:} 
		Similarly, using \eqref{L transformation} on \eqref{EOM of Yang-Mills}, we get
		\bes{}
		\begin{eqnarray}&&
		\delta_{L^{(n)}}\eqref{EOM-1}=(t^{n+1}\partial_{t}+(n+1)t^{n}x^{k}\partial_{k}+3(n+1)t^{n})\eqref{EOM-1}\non\\&&\hspace{2.2cm}
		+n(n+1)t^{n-1}x^{k}\eqref{EOM3} +n(n+1)(n-1)t^{n-2}x^{k}D_{k}\phi^{a}\non\\&&\hspace{6cm}+n(n+1)(n-1)t^{n-2}\phi^{a},\\&&
		\delta_{L^{(n)}}\eqref{EOM2}=(t^{n+1}\partial_{t}+(n+1)t^{n}x^{k}\partial_{k}+3(n+1)t^{n})\eqref{EOM2}=0,\\&&
		\delta_{L^{(n)}}\eqref{EOM3}=(t^{n+1}\partial_{t}+(n+1)t^{n}x^{k}\partial_{k}+3(n+1)t^{n})\eqref{EOM3}\non\\&&
		\hspace{5cm}+n(n+1)t^{n-1}x^{k}\eqref{EOM2}=0.
		\end{eqnarray}\ees
		We see that the (\ref{EOM2}, \ref{EOM3}) are invariant under $L^{(n)}$ for all $n$. However, \eqref{EOM-1} is invariant only under $L^{(-1,0,1)}$. \\\\
	In conclusion, the equations are completely invariant under $M^{(n)}_{k}$ but not invariant under $L^{(n)}$ for all $n$. They come out to be invariant under the global part of $L^{(-1,0,1)}\equiv(H,D,K)$.
\\\\
\textbf{\underline{Symmetries of equations when $\phi^{a} =0$:}} 
		When we take $\phi^a=0$ in \eqref{EOM of Yang-Mills}, we get the equations given in \eqref{phi0}. We will now look for the symmetries of these equations under infinite extension of GCA. To find the fields variation for this case, we have to put $\phi^{a}=0$ in the transformations (\ref{mnk}, \ref{L transformation}). The invariance of \eqref{phi0} under $L^{(n)}$ is given by
		\bes{}
		\begin{eqnarray}&&
		\delta_{L^{(n)}} (D_{i}E^{ia})=\Big[t^{n+1}\partial_{t}+(n+1)t^{n}x^{k}\partial_{k}+3(n+1)t^{n}\Big](D_{i}E^{ia})\non
		\\&&\hspace{5.7cm}+n(n+1)t^{n-1}x^{k}(D^{i}W_{ki}^{a})=0,\\&&
		\delta_{L^{(n)}} (D^{i}W_{ij}^{a})=\Big[t^{n+1}\partial_{t}+(n+1)t^{n}x^{k}\partial_{k}+3(n+1)t^{n}\Big](D^{i}W_{ij}^{a})=0.
		\end{eqnarray}\ees
		Similarly, under $M^{(n)}_k$
		\begin{eqnarray}&&
		\delta_{M^{(n)}_{k}}(D_{i}E^{ia})=0,~~	\delta_{M^{(n)}_{k}}(D^{i}W_{ij}^{a})=0.
		\end{eqnarray}
		These equations are invariant under full infinite dimensional GCA.

\medskip

So, in conclusion, the symmetry analysis of Galilean Yang-Mills in $d=4$ obtained by null reducing the relativistic Yang-Mills action in $d=5$ gives us  
\begin{itemize}
\item{\em{At the level of action}}: Invariance under finite GCA + infinite dimensional boosts.
\item{\em{At the level of EOM}}:  Invariance under finite GCA + infinite dimensional boosts.
\item{\em{At the level of EOM, with fields turned off}}: Invariance under infinite GCA. 
\end{itemize}	
An interesting point to re-emphasise is that the EOM even with the fields $\phi^a$ turned off are different from the ones obtained by taking limits as was described earlier in the section. It is possible that one needs to consider scalings of the gauge coupling $g$ in order to obtain these results{\footnote{A similar phenomenon was observed when constructing the action of Carrollian scalar electrodynamics in \cite{Bagchi:2019clu}.}}.

	\subsection{Spacetime symmetries of Galilean Yang-Mills theory in $d=3$}\label{gym3d}
	We will now focus on symmetries of the Galilean Yang-Mills action in three dimensions. We will show that the Galilean Yang-Mills action is invariant under Schr\"{o}dinger symmetries. The symmetry analysis for this theory under the Galilean transformations $(H, P_i, G_i, J_{ij})$ is very similar to the $d=4$ case, as discussed in the previous section. The following straightforward yet crucial analysis reveals the infinite number of $Y^{(n)}$ generators to be symmetries of theory.
	
\subsubsection{Symmetries at the level of action}	
	Under scale transformation \eqref{stsa}, the action transforms as
	\begin{equation}
	\delta_{\tilde{D}} S_{GYM}=\int dt\,d^2 x\, \Big(\delta_{\tilde{D}}\mathcal{L}_{GYM}\Big)
	= \int dtd^2x\big[\partial_{t}(2t\mathcal{L}_{GYM})+\partial_{k}( x^{k}\mathcal{L}_{GYM})\big],
	\end{equation}
	which is a total derivative, implying the invariance of the action under dilatation.
	Under the special conformal transformation \eqref{sctsa}, the action changes as
	\begin{equation}
	\delta_{\tilde{K}} S=\int dtd^2x\big[\p_{t}(2t^{2}\mathcal{L}_{GYM})+\p_{k}(2tx^{k}\mathcal{L}_{GYM})\big].
	\end{equation}
	Thus the action is invariant under special conformal transformation in $d=3$. We will now move on to the invariance of the action under the infinite extension of Schr\"{o}dinger algebra. Under $Z^{(n)}$  and $Y_{k}^{(n)}$  transformations, the fields transform as
	\begin{eqnarray}\label{3d L transformation Yang-Mills}&&
		\delta_{Z^{(n)}}\phi^{a}=2t^{n+1}\p_t \phi^{a}+(n+1)t^{n}(x^{k}\partial_{k}+\Delta_{\phi})\phi^{a}, \nonumber \\&&
		\delta_{Z^{(n)}}a_{i}^{a}=2t^{n+1}\p_t a_{i}^{a}+(n+1)t^{n}(x^{k}\partial_{k}+\Delta_{a_i})a_{i}^{a}-n(n+1)t^{n-1}x_{i}\phi^{a}, \nonumber \\&&
		\delta_{Z^{(n)}}a_{t}^{a}=2t^{n+1}\p_t {a}_{t}^{a}+(n+1)t^{n}(x^{k}\partial_{k}+\Delta_{a_t})a_{t}^{a}+n(n+1)t^{n-1}x^{k}a_{k}^{a}.
	\end{eqnarray}
	Similarly for $Y$, we have
	\begin{eqnarray}\label{3d M trans Yang-Mills}&&
		\delta_{Y^{(m)}_{k}}\phi^{a}=t^{m+\frac{1}{2}}\, \partial_{k}\phi^{a},~
		\delta_{Y^{(m)}_{k}}a_{i}^{a}=t^{m+\frac{1}{2}}\partial_{k}a_{i}^{a}-\big(m+\frac{1}{2}\big)t^{m-\frac{1}{2}}\phi^{a},\non\\&&
		\delta_{Y^{(m)}_{k}}a_{t}^{a}=t^{m+\frac{1}{2}}\partial_{k}a_{t}^{a}+\big(m+\frac{1}{2}\big)t^{m-\frac{1}{2}}a_{k}^{a}.
	\end{eqnarray}
Under $Z$ and $Y$ transformations above, the changes in the action are
	\begin{eqnarray}&&\hspace{-1cm}
	\delta_{Z^{(n)}} S_{GYM}=\int dtd^2x \Big[\partial_{t}\Big(2t^{n+1}\mathcal{L}_{GYM}\Big)+\partial_{k}\Big(\,(n+1)t^{n}x^{k}\mathcal{L}_{GYM}\Big)\non\\&&\hspace{3cm}-n(n+1)(n-1)\,t^{n-2}(\phi^{a})^2\Big], \\
	&&\hspace{-1cm}	\delta_{Y^{(n)}_{k}} S_{GYM} =\int dt\,d^2x\,\partial_{k}\Big[- \Big(t^{n+\frac{1}{2}}\,\mathcal{L}_{GYM}-\frac{1}{2}(n-\frac{1}{2})(n+\frac{1}{2})t^{n-\frac{3}{2}}(\phi^a)^2\Big)\Big].
	\end{eqnarray}
In conclusion, the action comes out to be invariant under $Z^{(-1,0,1)}\equiv(H,D,K)$ and completely invariant under $Y^{(n)}_{k}$ transformations.

\subsubsection{Symmetries at the level of EOM}
We will focus on the invariance of the EOM under the infinite dimensional Schr\"odinger algebra. 
\\\\
\textbf{\underline{$Y_{k}^{(n)}$ transformations}:} The invariance of EOM are given as below
\bes{}
\begin{eqnarray}&&
\delta_{Y^{(n)}_{k}} \eqref{EOM-1} = -t^{n+\frac{1}{2}}\partial_{k}\eqref{EOM-1}-(n+\frac{1}{2})t^{n-\frac{1}{2}}\eqref{EOM3}=0,\\&&
\delta_{Y^{(n)}_{k}} \eqref{EOM3} = -t^{n+\frac{1}{2}}\partial_{k}\eqref{EOM3}-(n+\frac{1}{2})t^{n-\frac{1}{2}}\delta_{ik}\eqref{EOM2}=0,\\&&
\delta_{Y^{(n)}_{k}} \eqref{EOM2} = -t^{n+\frac{1}{2}}\partial_{k}\eqref{EOM2}=0.
\end{eqnarray}
\ees
\textbf{\underline{$Z^{(n)}$ transformations}:} 
Under \eqref{3d L transformation Yang-Mills}, the equatons \eqref{EOM of Yang-Mills} transforms as
\bes{}
\begin{eqnarray}&&
\delta_{Z^{(n)}}\eqref{EOM-1}=(2t^{n+1}\partial_{t}+(n+1)t^{n}x^{k}\partial_{k}+4(n+1)t^{n})\eqref{EOM-1}\non\\&&\hspace{2.2cm}
+n(n+1)t^{n-1}x^{k}\eqref{EOM3}-2n(n+1)(n-1)t^{n-2}\phi^{a},\\&&
\delta_{Z^{(n)}}\eqref{EOM2}=(2t^{n+1}\partial_{t}+(n+1)t^{n}x^{k}\partial_{k}+2(n+1)t^{n})\eqref{EOM2}=0,\\&&
\delta_{Z^{(n)}}\eqref{EOM3}=(2t^{n+1}\partial_{t}+(n+1)t^{n}x^{k}\partial_{k}+3(n+1)t^{n})\eqref{EOM3}\non\\&&
\hspace{5cm}-n(n+1)t^{n-1}x^{k}\eqref{EOM2}=0.
\end{eqnarray}\ees
We see that the equations \eqref{EOM of Yang-Mills} are invariant under $Z^{(-1,0,1)}$ and $Y_{k}^{(n)}$ for all $n$. 
Now for the case with $\phi^a=0$. The invariance of \eqref{phi0} under $Z^{(n)}$ is given by
\bes{}
\begin{eqnarray}&&
	\delta_{Z^{(n)}} (D_{i}E^{ia})=\Big[2t^{n+1}\partial_{t}+(n+1)t^{n}x^{k}\partial_{k}+4(n+1)t^{n}\Big](D_{i}E^{ia})\non
	\\&&\hspace{5.7cm}-n(n+1)t^{n-1}x^{k}(D^{i}W_{ki}^{a})=0,\\&&
	\delta_{Z^{(n)}} (D^{i}W_{ij}^{a})=\Big[2t^{n+1}\partial_{t}+(n+1)t^{n}x^{k}\partial_{k}+2(n+1)t^{n}\Big](D^{i}W_{ij}^{a})=0.
\end{eqnarray}\ees
Similarly, under $Y^{(n)}_k$
\begin{eqnarray}&&
	\delta_{Y^{(n)}_{k}}(D_{i}E^{ia})=0,~~	\delta_{Y^{(n)}_{k}}(D^{i}W_{ij}^{a})=0.
\end{eqnarray}
We see that the equations \eqref{phi0} are invariant under the infinite extension of Schr$\ddot{o}$dinger symmetry.

\medskip

Again to summarise, the symmetry analysis of Galilean Yang-Mills in $d=3$ obtained by null reducing the relativistic Yang-Mills action in $d=4$ gives us  
\begin{itemize}
\item{\em{At the level of action}}: Invariance under finite Schr{\"o}dinger algebra + infinite dimensional boosts.
\item{\em{At the level of EOM}}:  Invariance under finite Schr{\"o}dinger algebra + infinite dimensional boosts.
\item{\em{At the level of EOM, with field turned off}}: Invariance under infinite Schr{\"o}dinger-Virasoro algebra. 
\end{itemize}

\subsection{Scale transformation with generic $z$}
Dynamical exponent $z$ is defined as the relative scaling of space and time under the Dilatation operator:
\be{}
D: \quad t \to \lambda^z t, \quad x \to \lambda x. 
\ee
For the case of the Schr{\"o}dinger algebra $z=2$ and, rather interestingly, for the GCA $z=1$. Usually, $z=1$ indicates a relativistic theory, but GCFTs are examples of non-relativistic QFTs with $z=1$. 

We now discuss whether scale transformations with generic dynamical exponent $z$ can be a symmetry at the level of Lagrangian \eqref{ngym} in $d$ spacetime dimensions. To understand it, we first write down the scale transformation of different fields $\Phi^a = (\phi^{a},\,a_{t}^{a},\,a_{i}^{a})$, given by
	\be{std}
	\delta_{\bar{D}}\Phi^a=-(zt\partial_{t}+x^{k}\partial_{k}+\Delta_{\Phi}) \Phi^a.
	\ee
	We will now look at the transformation of different covariant terms \eqref{quan} under \eqref{std}:
	\begin{eqnarray}&&\non
	\hspace{-1.7cm}\delta_{\bar{D}}(D_{t}\phi^{a})=\delta_{\bar{D}}(\partial_{t}\phi^{a}-gf^{abc}\phi^{b}a_{t}^{c})\\&&
	= -(zt\partial_{t}+x^{k}\partial_{k})D_{t}\phi^{a}-(\Delta_{\phi}+z)\partial_{t}\phi^{a}+(\Delta_{\phi}+\Delta_{a_{t}})gf^{abc}\phi^{b}a_{t}^{c}		
	\end{eqnarray}
If we demand that $(D_{t}\phi^{a})$ transforms as a primary, we find 
\be{}
\Delta_{\phi}+z=\Delta_{\phi}+\Delta_{a_{t}} \Rightarrow \Delta_{a_{t}}=z.
\ee 
Similarly, demanding the same for $D_{i}\phi^{a}, W_{ij}^{a}, E^{ia}$, we get
\be{}\Delta_{\phi}+1=\Delta_{\phi}+\Delta_{a_{i}} \implies \Delta_{a_{i}}=1\ee
	Finally, we will now look at the change in the action. The remaining terms are given by
	\bea{}&&\hspace{-1.5cm}\delta_{\tilde{D}} S_{GYM}=\int dt\,d^{d-1} x\, \Big\{\Big[\frac{d-1-z}{2}-\Delta_{\phi}\Big](D_{t}\phi^{a})^2-(\Delta_{\phi}+\Delta_{a_{i}}-d+2)D_{i}\phi^{a}E^{ia}\non\\&&\hspace{6cm}-\Big[\frac{z+(d-1)}{4}-\frac{\Delta_{a_{i}}+1}{2}\Big]W^{ija}W_{ij}^{a}\Big\}.
	\eea
The conditions we get from above are
	\begin{eqnarray}
	\Delta_{a_{i}}=1,~ \Delta_{a_{t}}=z, ~ d-1-z=2\Delta_{\phi},~ \Delta_{\phi}=d-3.
	\end{eqnarray}
So, from these equations we arrive at 
\be{}
z+d=5.
\ee
The significance of this equation for $d>5$ is not clear to us. This condition should be satisfied to get the dilatation invariance of the Lagrangian \eqref{ngym}. Clearly, we get $z=1$ when $d=4$ and $z=2$ when $d=3$, which are both in agreement with our analysis in the paper so far.

\newpage
	
\section{Noether charges and phase space analysis}
The analysis above clearly reflects that the infinite dimensional Abelian ideal along with 3 generators of the $SL(2)$ subalgebra for both the GCA and the Schrodinger algebra, respectively for $d=4$ and $d=3$ are symmetries of the theory. In the following we will investigate the dynamical realization of these symmetries in terms of Noether charges. However, for both the cases, we also noticed that the infinite dimensional enhancement of the $SL(2)$ part, i.e. the Witt algebra generates on-shell symmetries, at the level of equations of motion, when a particular field is turned off. This is a reflection of the fact that these are weak symmetries in the sense elaborated in \cite{Beisert:2017pnr,Beisert:2018zxs,Bagchi:2019clu,Banerjee:2020qjj}. To investigate the roles of these later ones, we look at their realizations on phase space, where they generate Hamiltonian vector fields.
	\subsection{Galilean Yang-Mills theory in $d=4$}
We will now examine the dynamical realization of the infinite dimensional GCA on the space of fields satisfying the EOM. The Lagrangian is invariant under $M^{(n)}_{i}$ and finite part of $L^{(n)}$, obviously giving rise to conserved quantities. In spite of the fact that $L^{(n)}, \, \forall n \in \mathbb{Z}$ are not symmetry generators of the theory \eqref{ngym}, we will see that they are the symmetries of the phase space.
	
In order to construct the conserved quantities, we make use of the pre-symplectic potential on the space of solutions, defined from the Lagrangian via:
	\bea{}\label{theta}
	\delta L= \int d^{d-1}x \:\p_t \underbrace{\Theta( \delta )}_{\mbox{(pre)-symplectic potential}}  ~~: \text{on-shell}
	\eea
for a generic variation $\delta \varphi$. The pre-symplectic potential arising from \eqref{ngym} is \bea{}\Theta (\delta)=D_{t}\phi^{a}(\delta\phi^{a})+D_{i}\phi^{a}(\delta a^{ia}).\eea
	Now, for a specific symmetry transformation  $\varphi \to \varphi +\delta_{\epsilon} \varphi$:
	\bea{}\label{alpha}
	\delta_{\e} L= \int d^{d-1}x \:\p_t \b  ( \delta_{\e})  ~~: \text{off-shell,}
	\eea
	for some function $\b$ in field space. Hence, comparing \eqref{theta} and \eqref{alpha}, we deduce that on-shell, the Noether charges $Q_{\e}$ are given by :
	\bea{}\label{inter}
	 Q_{\e} = \int d^{d-1}x  \left( \Theta(\delta_{\epsilon}) - \b  (\delta_{\e}) \right)
	\eea
	The Noether charges for GCA in $d=4$ can now be calculated using this technique. We will focus on the Noether charges for $L^{(-1,0,1)}=H,D,K$.
	\begin{eqnarray}&&\hspace{-1.3cm}
	Q_{H}=\int d^3x \,\Gamma\equiv H, \label{hamiltonmatch}\\&&\hspace{-1.3cm}
	Q_{D}=\int d^3x \,\Big[t\Gamma+(x^{k}\partial_{k}\phi^{a}+\phi^{a})D_{t}\phi^{a}+(x^{k}\partial_{k}a_{i}^{a}+a_{i}^{a})D_{i}\phi^{a}\Big],\\&&\hspace{-1.3cm}
	Q_{K}=\int d^3x\, \Big[t^2\Gamma+2t(x^{k}\partial_{k}\phi^{a}+\phi^{a})D_{t}\phi^{a}+(2tx^{k}\partial_{k}a_{i}^{a}+2ta_{i}^{a}-2x_{i}\phi^{a})D_{i}\phi^{a}+(\phi^{a})^{2}\Big],
	\end{eqnarray}
	where $$\Gamma=\frac{1}{2}D_{t}\phi^{a}D_{t}\phi^{a}+D_{i}\phi^{a}D_{i}a_{t}^{a}+\frac{1}{4}W^{ija}W_{ij}^{a}+gf^{abc}\phi^{b}a_{t}^{c}D_{t}\phi^{a}$$ and $H$ is the Hamiltonian of our system.
Interestingly, as in the Abelian case \cite{Banerjee:2019axy} the charges corresponding to the $M^{(n)}_{k}$ transformations vanish for all $n$:
	\be{}Q_{M}=\int d^{3}x \, \left(\Theta(\delta_{M^{(n)}_{k}})  - \beta(\delta_{M^{(n)}_{k}}) \right)=\int d^3 x \, \left( D_{t}\phi^{a}(\delta_{M^{(n)}_{k}}\phi^{a})+D_{i}\phi^{a}(\delta_{M^{(n)}_{k}} a^{ia})\right) = 0.\ee

	\subsection*{Pre-symplectic structure}
	In order to inspect the action of the $L^{(n)}$ generators on the phase space, we start with the pre-symplectic structure for Galilean Yang-Mills theory given by the exterior derivative of the pre-symplectic potential on the space of fields:
	\begin{eqnarray}\label{pre-symplectic potential}&&
	\Omega(\delta_{1},\delta_{2})= \int d^3 x \,  \left( \delta_2 \Theta (\delta_1) - \delta_1 \Theta (\delta_2) \right) \non \\&&
=	\int d^3 x\,[D_{t}\delta_{2}\phi^{a}\delta_{1}\phi^{a}+D_{i}\delta_{2}\phi^{a}\delta_{1}a^{ia}-D_{t}\delta_{1}\phi^{a}\delta_{2}\phi^{a}-D_{i}\delta_{1}\phi^{a}\delta_{2}a^{ia}\non\\&&
	\hspace{1.9cm}-gf^{abc}(\phi^{b}\delta_{2}a_{t}^{c}\delta_{1}\phi^{a}+\phi^{b}\delta_{2}a_{i}^{c}\delta_{1}a^{ia}-\phi^{b}\delta_{1}a_{t}^{c}\delta_{2}\phi^{a}-\phi^{b}\delta_{1}a_{i}^{c}\delta_{2}a^{ia})].
	\end{eqnarray}
As expected in any gauge theory, the gauge transformation \eqref{gtf} gives a degenerate direction in space of solutions:
	\begin{eqnarray}
	\Omega(\delta_{\alpha^{a}},\delta)=0
	\end{eqnarray}
	 for any arbitrary gauge parameter $\alpha^{a}$.
	
	\medskip
	
	\noindent \textbf{\underline{$M_{k}^{(n)}$ and $L^{(n)}$ transformations:}}
   Consistent with the finding that conserved quantities associated with the $M_{k}^{(n)}$ transformations \eqref{mnk} are all 0, we observe that their corresponding Hamiltonian functions are trivially zero as well:
	\begin{eqnarray}
	\Omega(\delta_{M_{k}^{(n)}},\delta)=0.
	\end{eqnarray}
Curiously, $L^{(n)}$ transformations for all $n$ are Hamiltonian vector fields, giving rise to non-trivial generators of canonical transformations, despite the fact that only the $n=0, \pm 1$ modes are symmetries of the theory:
	\begin{eqnarray}&&\hspace{-.5cm}
	\Omega(\delta_{L^{(n)}},\delta)=\delta \int d^{3}x\,\Big[t^{n+1} \Big\{\frac{1}{2}D_{t}\phi^{a}D_{t}\phi^{a}+\frac{1}{4}W^{ija}W_{ija}\Big\}-(n+1)t^{n}\Big\{2(a^{ia}D_{i}\phi^{a}+\phi^{a}D_{t}\phi^{a})\non\\&&\hspace{2.5cm}
	+x^{k}(\phi^{a}\partial_{k}D_{t}\phi^{a}+a_{i}^{a}\partial_{k}D^{i}\phi^{a})\Big\}+n(n+1)t^{n-1}(\phi^{a})^{2}\Big]=\delta Q[f(t)].
	\end{eqnarray}

	\medskip
	
	\noindent \textbf{\underline{Algebra of the Hamiltonian functions:}}
	One of the most important questions in the phase space analysis of GCA in context of Galilean Yang Mills theory is whether the Witt sub-algebra of GCA generated by the Hamiltonian vector fields $\delta_{L^{(n)}}$ gets centrally extended or not at the level of the corresponding Hamiltonian functions. The answer turns out to be affirmative:
	\begin{eqnarray}
		\Omega(\delta_{L^{(n)}},\delta_{L^{(m)}})=(n-m)Q[L^{(n+m)}] + K_{L^{(n)},L^{(m)}}
	\end{eqnarray}
	where $K_{L^{(n)},L^{(m)}}  =\big[n(n^{2}-1)-m(m^{2}-1)\big]t^{n+m-1}\int d^{3}x (\phi^{a})^{2} $ is a state dependent central extension, routinely observed in asymptotic symmetries of gravitational theories \cite{Barnich:2011mi} and satisfies the 2 co-cycle condition:
	\begin{equation} \label{cohom}
	K_{[L^{(n)},L^{(m)}],L^{(r)}} +  K_{[L^{(m)},L^{(r)}],L^{(n)}}+K_{[L^{(r)},L^{(n)}],L^{(m)}} = 0.  
	\end{equation}
The same feature was observed in the in the Abelian case \cite{Banerjee:2019axy} as well. The realization of the commutation relation $[L^{n} , M^{(m)}_{k}]=(n-m) M^{(n+m)}_{k}$ is trivial as the Hamiltonian functions corresponding to $ M^{(m)}_{k}$ are all zero.

	\subsection{Galilean Yang-Mills theory in $d=3$}
 Since the Lagrangian in $d=3$ has the same form	as that of $d=4$, the construction for Noether charges follow exactly the same route. Before going into the details of some of the expressions for charges and the algebra, it is important to point out that everything that we construct in this sub-section also holds for the $U(1)$ case, just by turning off the structure constants. So our conclusions for this sub-section equally applies to Galilean Electrodynamics in $d=3$. 
 
Conserved quantities corresponding to the $Z^{(n)}, \, (n = 0, \pm 1)$ symmetry generators \eqref{3d L transformation Yang-Mills} of the Schr\"odinger algebra are as follows.
	\begin{eqnarray}&&\hspace{-1cm}
	Q_{H}=\int d^2x \,\mathcal{H}\equiv \text{Hamiltonian}, \label{hamiltonmatch}\\&&\hspace{-1cm}
	Q_{\tilde{D}}=\int d^2x \,\Big[t\mathcal{H}+x^{k}\partial_{k}\phi^{a}D_{t}\phi^{a}+(x^{k}\partial_{k}a_{i}^{a}+a_{i}^{a})D_{i}\phi^{a}\Big],\\&&\hspace{-1cm}
	Q_{\tilde{K}}=\int d^2x\, \Big[t^2\mathcal{H}+2tx^{k}\partial_{k}\phi^{a}D_{t}\phi^{a}+(2tx^{k}\partial_{k}a_{i}^{a}+2ta_{i}^{a}-2x_{i}\phi^{a})D_{i}\phi^{a}\Big],
	\end{eqnarray}
	where $\mathcal{H}=2\Big[\frac{1}{2}D_{t}\phi^{a}D_{t}\phi^{a}+D_{i}\phi^{a}D_{i}a_{t}^{a}+\frac{1}{4}W^{ija}W_{ij}^{a}+gf^{abc}\phi^{b}a_{t}^{c}D_{t}\phi^{a}\Big]$.
	However the charges corresponding to the $Y^{(n)}_{k}$ transformations vanish trivially for all $n$, analogous to the role of the $M^{(n)}$ generators in the 4 dimensional counterpart:
	\be{}Q_{Y}=\int d^{3}x \, \left(\Theta(\delta_{Y^{(n)}_{k}})  - \beta(\delta_{Y^{(n)}_{k}}) \right)=\int d^3 x \, \left( D_{t}\phi^{a}(\delta_{Y^{(n)}_{k}}\phi^{a})+D_{i}\phi^{a}(\delta_{Y^{(n)}_{k}} a^{ia})\right) = 0.\ee
	\subsubsection*{Hamiltonian functions}
	As expected from the symmetry analysis above, we notice that the transformations $\delta_{Y_{k}^{(n)}}$ are Hamiltonian vector fields on the space of solutions, but with vanishing dynamical generators:
	\begin{equation}
	\Omega({\delta_{Y_{k}^{(n)}},\delta})=0.
	\end{equation}
On the other hand, all the infinite modes of the $Z^{(n)}$ generators give rise integrable Hamiltonian functions on the space of solutions. Substituting these expressions in \eqref{pre-symplectic potential}, we have
	\bea{}&&\label{}\hspace{-.5cm}
	\Omega(\delta_{Z^{(n)}},\delta)=
	\delta \int d^{2}x\,\Big[2t^{n+1} \Big\{\frac{1}{2}D_{t}\phi^{a}D_{t}\phi^{a}+\frac{1}{4}W^{ija}W_{ija}\Big\}-(n+1)t^{n}\Big\{(a^{ia}D_{i}\phi^{a}+2\phi^{a}D_{t}\phi^{a})\non\\&&\hspace{2.8cm}
	+x^{k}(\phi^{a}\partial_{k}D_{t}\phi^{a}+a_{i}^{a}\partial_{k}D^{i}\phi^{a})\Big\}+(n+1)nt^{n-1}(\phi^{a})^{2}\Big]=\delta Q[Z_{n}].
	\eea
For the phase space realization of the Schr\"odinger algebra \eqref{schrod}, we readily notice that the central mass term drops off, since the Hamiltonian functions corresponding to $Y$ generators themselves vanish. For the very same reason, the brackets $[ Z^{(n)},Y_{i}^{(m)}] $ trivialize at the level of Hamiltonian functions:
	\begin{equation}
	\Omega(\delta_{Z^{(n)}},\delta_{Y_{i}^{(m)}})=0.
	\end{equation}
	However the Witt subalgebra of the Schr\"odinger algebra generated by $Z^{(n)}$  gets centrally extended on the space of solutions:
	\begin{equation} \label{d3vir}
	\Omega(\delta_{Z^{(n)}},\delta_{Z^{(m)}})=(n-m)Q[Z^{(n+m)}]+2\big[n(n^{2}-1)-m(m^{2}-1)\big]t^{n+m-1}\int d^{3}x (\phi^{a})^{2}.
	\end{equation}
	As in the Witt subalgebra of GCA, the second term in \eqref{d3vir} is also a state dependent central term satisfying the 2 co-cycle condition \eqref{cohom}. So indeed, even for Galilean gauge theories in 3d (both electrodynamics and Yang-Mills), the entire infinite dimensional Virasoro sub-algebra of the Schr{\"o}dinger-Virasoro algebra is realised in phase space, along with the central extensions.  	This is rather remarkable given that this same sub-algebra did not turn out to be symmetries of the action or the subsequent EOM. 	
\bigskip	

\newpage

	\section{Quantum Aspects: Propagators and vertices}	
	
	In earlier sections, we focused on classical aspects of Galilean electrodynamics and Galilean Yang-Mills theories. Now we would to like to look at the quantization of these theories. In this paper, we take the initial steps to our goal. 
		
	\medskip
	
	\noindent A detailed study of the quantum theory of GED coupled to a scalar field in $3$-dimensions was done in \cite{Chapman:2020vtn}. For our purposes, where the additional scalar field is absent, we only review here the expression for the propagators of the gauge fields of GED:
	\begin{eqnarray}\label{Abelian propagator}
		\langle \mathcal{A}_I(k)\mathcal{A}_J(-k)\rangle \equiv D_{IJ}(k) = -\frac{i}{\vec{k}^2}\begin{pmatrix}
			0 &\quad 1 &\quad 0 \\
			1 &\quad  -(1-\xi)\frac{\omega^2}{\vec{k}^2} &\quad  (1-\xi)\frac{\omega k_j}{\vec{k}^2} \\
			0 &\quad  (1-\xi)\frac{\omega k_i}{\vec{k}^2} &\quad  \delta_{ij} - (1-\xi)\frac{k^i k^j}{\vec{k}^2}
		\end{pmatrix},
	\end{eqnarray}
	where  $k = (\omega,\vec{k})$, $\mathcal{A}_I = (\phi,a_t,a_i)$ and $i=\{1,2\}$. (Performing a similar analysis in $4$-dimensions, we get the same expression \eqref{Abelian propagator} for gauge field propagators in $4$-dimensions with $i=\{1,2,3\}$.) This will be useful when we try to find the propagators for the Galilean Yang-Mills theory. 
	\begin{figure}[h]
		\centering
		\includegraphics[width=0.3\linewidth]{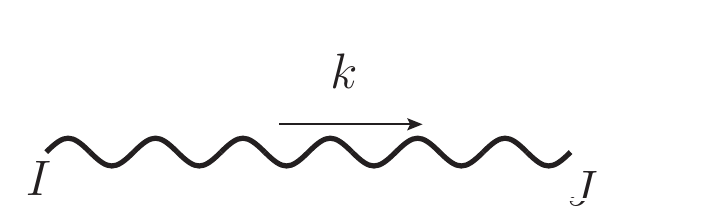}
		\caption{GED propagator $\langle \mathcal{A}_I\mathcal{A}_J\rangle$.}
		\label{fig:abp}
	\end{figure}
	
	\medskip
	
	\noindent The construction of the full quantum theory of Galilean Yang-Mills theory is quite non-trivial. In this section, we begin by obtaining tree level propagators and vertices, and write Feynman rules for Galilean Yang Mills theory. We hope to return to more details of the full quantum mechanical theory in future work.

Let us first recall the total gauge-fixed Lagrangian of the relativistic Yang-Mills theory including the gauge fixing and ghost terms:
	\begin{eqnarray}\label{relativisticfulllagrangian}
		\mathcal{L}=-\frac{1}{4}F^{\tilde{\mu}\tilde{\nu} a}F^{a}_{\tilde{\mu}\tilde{\nu}}-\frac{1}{2\xi}\big(\partial^{\tilde{\mu}}A_{\tilde{\mu}}\big)^{2}-\partial^{\tilde{\mu}}\bar{c}^{a}D_{\tilde{\mu}}c^{a},
	\end{eqnarray}
	where $\xi$ is the gauge fixing parameter, $(c^a,\bar{c}^a)$ are ghost fields and $D_{\mu}c^{a}=\partial_{\mu}c^{a}-gf^{abc}A_{\mu}^{b}c^{c}$. Performing null reduction along $u$-direction, we get the total Lagrangian density, including the gauge fixing term and the ghost term, for Galilean Yang-Mills theory in four dimensions
	\begin{eqnarray}
	\tilde{\mathcal{L}}_{GYM} &=& \frac{1}{2}D_{t}\phi^{a}D_{t}\phi^{a}+D_{i}\phi^{a}E^{ia}-\frac{1}{4}W^{ija}W_{ij}^{a} \nonumber \\
	&& -\frac{1}{2\xi}(\partial_{t}\phi^{a}+\partial^{i}a_{i}^{a})^{2}-gf^{abc}\partial_{t}\bar{c}^{a}\phi^{b}c^{c} -\delta^{ij}\partial_{i}\bar{c}^{a}(D_jc)^a. \label{GYM-Lagrangian-4d}
	\end{eqnarray}
	From this Lagrangian density, we have the kinetic terms as
	\begin{eqnarray}
	\mathcal{L}_{kin} &=& \frac{1}{2}\partial_{t}\phi^{a}\partial_{t}\phi^{a} + \delta^{ij}\partial_{i}\phi^{a}(\partial_{t}a^{a}_j-\partial_{j}a_{t}^{a}) - \frac{1}{2}\delta^{ik}\delta^{jl}(\partial_{i}a_j^{a}\partial_{k}a_{l}^{a} - \partial_{i}a_j^{a}\partial_{l}a_{k}^{a}) \nonumber \\
	&& -\frac{1}{2\xi}(\partial_{t}\phi^{a}\partial_{t}\phi^{a}+2\partial_{t}\phi^{a}\partial^{i}a_{i}^{a} + \delta^{ij}\delta^{kl}\partial_{i}a_{j}^{a}\partial_{k}a_{l}^{a}) - \delta^{ij}\partial_{i}\bar{c}^{a}\partial_{j}c^{a}.
	\end{eqnarray}
	In order to get propagators from this kinetic part of the Lagrangian, let us first introduce Fourier transformation to momentum space
	\begin{equation}\label{GYM-fourier-transform}
	\Phi^a(t,\vec{x}) = \int \frac{d\omega}{2\pi}\frac{d^3\vec{k}}{(2\pi)^3}e^{-i\omega t}e^{i\vec{k}\cdot\vec{x}}\tilde{\Phi}^a(\omega,\vec{k}),
	\end{equation}
	where $\Phi^a = (\phi^a, a_t^a, a_i^a, c^a, \bar{c}^a)$, and delta functions
	\begin{equation}\label{GYM-delta-functions}
	\int\frac{d\omega}{2\pi}e^{-i\omega t} = \delta(\omega), \quad \int \frac{d^3\vec{x}}{(2\pi)^3} e^{i\vec{k}\cdot\vec{x}} = \delta^{(3)}(\vec{k}).
	\end{equation}
	We also introduce the notation, $k = (\omega,\vec{k})$ and $\mathcal{A}_I^a = (\phi^a,a_t^a,a_i^a)$. Taking Fourier transformation and using delta functions, the kinetic part of the action becomes
	\begin{equation}\label{GYM-kin-action-4d}
	\mathcal{S}_{kin} = \int \frac{d\omega d^3\vec{k}}{(2\pi)^4}\Big(\frac{1}{2}\mathcal{A}_I^a(k) d^{IJab}\mathcal{A}_J^b(-k) +\bar{c}^a(k)\big(-\vec{k}^2\big)c^a(-k) \Big),
	\end{equation}
	where
	\begin{eqnarray}
	d^{IJab}(k) = \begin{pmatrix}
	\Big(1-\frac{1}{\xi}\Big)\omega^2 &\quad  -\vec{k}^2 &\quad -\Big(1-\frac{1}{\xi}\Big)\omega k^j \\
	-\vec{k}^2 &\quad 0 &\quad 0 \\
	-\Big(1-\frac{1}{\xi}\Big)\omega k^i &\quad 0 &\quad -\delta^{ij}\vec{k}^2 + \Big(1-\frac{1}{\xi}\Big)k^i k^j
	\end{pmatrix}.
	\end{eqnarray}
	Then from the inverse of $d^{IJab}$, we get the propagators for the fields $\mathcal{A}_I^a$ as
	\begin{eqnarray}\label{Yang-Mills propagator}
	\langle A_I^a(k)A_J^b(-k)\rangle \equiv D^{ab}_{IJ}(k) = -\frac{i\delta^{ab}}{\vec{k}^2}\begin{pmatrix}
	0 &\quad 1 &\quad 0 \\
	1 &\quad  -(1-\xi)\frac{\omega^2}{\vec{k}^2} &\quad  (1-\xi)\frac{\omega k_j}{\vec{k}^2} \\
	0 &\quad  (1-\xi)\frac{\omega k_i}{\vec{k}^2} &\quad  \delta_{ij} - (1-\xi)\frac{k^i k^j}{\vec{k}^2}
	\end{pmatrix}.
	\end{eqnarray}
	The inverse of the coefficient of $\bar{c}c$ in \eqref{GYM-kin-action-4d} gives the propagator for ghost fields as
	\begin{eqnarray}
	\langle \bar{c}^{a}(k)c^{b}(-k)\rangle =\frac{i\delta^{ab}}{\vec{k}^{2}}.
	\end{eqnarray}
	We see that the denominators of propagators above depend only on spatial momentum and are independent of frequency implying instantaneous propagation of field excitations. In other words, there are no local degrees of freedom in the theory. This is consistent with the fact that in spite of presence of the infinite number of $M{(n)}$ symmetry generators, they have no non-trivial dynamical realization.
	\begin{figure}[h]
		\centering
		\subfigure[Gauge field propagator $\langle \mathcal{A}^a_I\mathcal{A}^b_J\rangle$]{
			\label{fig:first}
			\includegraphics[height=1.5cm]{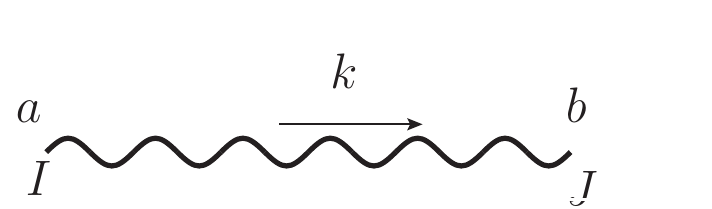}}
		\qquad
		\subfigure[Ghost propagator]{
			\label{fig:second}
			\includegraphics[height=1.5cm]{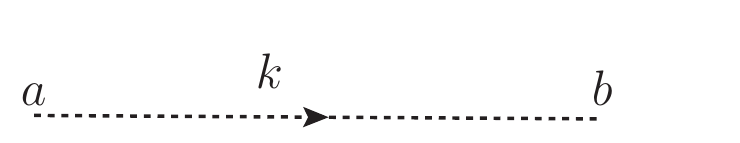}}
		\caption{Two-point Propagators}
	\end{figure}
		\\
	In order to obtain vertices, let us write the interaction part of the Lagrangian density \eqref{GYM-Lagrangian-4d}:
	\begin{eqnarray}
	\mathcal{L}_{int} &=& gf^{abc}\Big( \partial_{t}\phi^{a} a_{t}^{b}\phi^{c} + \delta^{ij}\partial_{i}\phi^{a}a_{t}^{b}a_j^{c} - \delta^{ij}a_i^b\phi^c\partial_j a_t^a + \delta^{ij}a_i^b\phi^c\partial_t a_j^a \nonumber \\
	&& - \delta^{ik}\delta^{jl}\partial_i a_j^a a_k^b a_l^c -\partial_t\bar{c}^a\phi^b c^c -\delta^{ij}\partial_i\bar{c}^a a_j^b c^c\Big) \nonumber \\
	&& + g^2 f^{abc}f^{ade}\Big(\frac{1}{2}a_t^b\phi^c a_t^d\phi^e + \phi^c a_t^d a_i^b a_j^e\delta^{ij} -\frac{1}{4}\delta^{ik}\delta^{jl}a_i^b a_j^c a_k^d a_l^e \Big).
	\end{eqnarray}
	Transforming to momentum space and using delta functions, we can write the $3$ field interaction terms in the action as
	\begin{eqnarray}
	\mathcal{S}^{(3)}_{int} &=& \int \frac{1}{(2\pi)^{12}}{\displaystyle{\prod_{i=1}^{3}d\omega_i d^3\vec{k}_i}}\,(2\pi)^4\delta(\omega_1 + \omega_2 + \omega_3) \delta^{(3)}(\vec{k}_1 + \vec{k}_2 + \vec{k}_3) g f^{abc} \times \nonumber \\
	&& \quad \Big[\frac{i}{2}(\omega_1 - \omega_2)\phi^a(k_1)\phi^b(k_2)a_t^c(k_3) + i\delta^{ij}(k_1 -k_2)_i \phi^a(k_1)a_t^b(k_2)a_j^c(k_3) \nonumber\\
	&&\quad + \frac{i}{2}(\omega_3 - \omega_2)\phi^a(k_1)a_i^b(k_2)a_j^c(k_3)\delta^{ij} \nonumber \\
	&& \quad  + \frac{i}{6}\big((k_1 - k_2)_i \delta^{il}\delta^{jk} + (k_2 - k_3)_i \delta^{ij}\delta^{lk} + (k_3 - k_1)_i \delta^{ik}\delta^{jl} \big)a_j^a(k_1)a_k^b(k_2)a_l^c(k_3)\nonumber\\
	&&\quad -i\omega_2 \phi^a(k_1)\bar{c}^b(k_2)c^c(k_3) + i\delta^{ij}k_{2j}a_i^a(k_1)\bar{c}^b(k_2)c^c(k_3) \Big],
	\end{eqnarray}
	where ${\displaystyle \prod_{i=1}^{n}d\omega_i d^3\vec{k}_i}=d\omega_1 d^3\vec{k}_1...d\omega_n d^3\vec{k}_n$. From this expression for the action in momentum space, we can write the $3$-point vertices as
	\begin{eqnarray}
		&& V_{3\,\phi\phi a_t}^{abc} = -g f^{abc}(\omega_1 - \omega_2), \quad V_{3\,\phi a_t a_i}^{abc\,i} = -g f^{abc}(k_1 - k_2)^i, \quad V_{3\,\phi a_i a_j}^{abc\, ij} = g f^{abc}(\omega_2 - \omega_3)\delta^{ij}, \nonumber\\
		&& V_{3\,a_i a_j a_k}^{abc\, ijk} = -g f^{abc}\big( (k_1 - k_2)^k\delta^{ij} + (k_2 - k_3)^i\delta^{jk} + (k_3 - k_1)^j\delta^{ik} \big), \nonumber \\
		&& V_{3\,\phi \bar{c} c}^{abc} = g f^{abc}\omega_2, \quad V_{3\,a_i \bar{c} c}^{abc} = - g f^{abc} k_2^i.
	\end{eqnarray}
	Similarly, transforming the $4$ field interaction terms in $\mathcal{S}_{int}$ to momentum space, we get
	\begin{eqnarray}
	\mathcal{S}^{(4)}_{int} &=& \int \frac{1}{(2\pi)^{16}}{\displaystyle{\prod_{i=1}^{4}d\omega_i d^3\vec{k}_i}}\,(2\pi)^4\delta(\omega_1 + \omega_2 + \omega_3 + \omega_4) \delta^{(3)}(\vec{k}_1 + \vec{k}_2 + \vec{k}_3 + \vec{k}_4) \times \nonumber \\
	&& \quad g^2 \Big[\frac{1}{4}\big(f^{abd}f^{ace} +f^{abe}f^{acd} \big)\phi^b(k_1)\phi^c(k_2)a_t^d(k_3)a_t^e(k_4) \nonumber \\
	&& \qquad - \big(f^{abd}f^{ace} +f^{abe}f^{acd} \big)\phi^b(k_1)a_t^c(k_2)a_i^d(k_3)a_j^e(k_4)\delta^{ij} \nonumber \\
	&&\qquad -\frac{1}{24}\big( f^{abc}f^{ade}(\delta^{ik}\delta^{jl} - \delta^{il}\delta^{jk}) + f^{abd}f^{ace}(\delta^{ij}\delta^{kl} - \delta^{il}\delta^{jk}) \nonumber \\
	&&\qquad  + f^{abe}f^{acd}(\delta^{ij}\delta^{kl} - \delta^{ik}\delta^{jl}) \big)a_i^b(k_1)a_j^c(k_2)a_k^d(k_3)a_l^e(k_4) \Big],
	\end{eqnarray}
	from which we can read of the $4$-point vertices 
	\begin{eqnarray}
	&& V_{4\,\phi\phi a_t a_t}^{bcde} = ig^2\big(f^{abd}f^{ace} + f^{abe}f^{acd}\big), \nonumber \\
	&& V_{4\,\phi a_t a_i a_j}^{bcde\,ij} = -ig^2\big(f^{abd}f^{ace} + f^{abe}f^{acd}\big)\delta^{ij}, \nonumber \\
	&& V_{4\,a_i a_j a_k a_l}^{bcde\,ijkl} = -4 ig^2\Big( f^{abc}f^{ade}(\delta^{ik}\delta^{jl} - \delta^{il}\delta^{jk}) + f^{abd}f^{ace}(\delta^{ij}\delta^{kl} - \delta^{il}\delta^{jk}) \nonumber \\
	&& \hspace{35mm} + f^{abe}f^{acd}(\delta^{ij}\delta^{kl} - \delta^{ik}\delta^{jl}) \Big).
	\end{eqnarray}
	\begin{figure}[h]
		\centering
		\subfigure[][$V_{3\,\phi\phi a_t}^{abc}$]{
			\label{fig:ex3-a}
			\includegraphics[height=3.3cm]{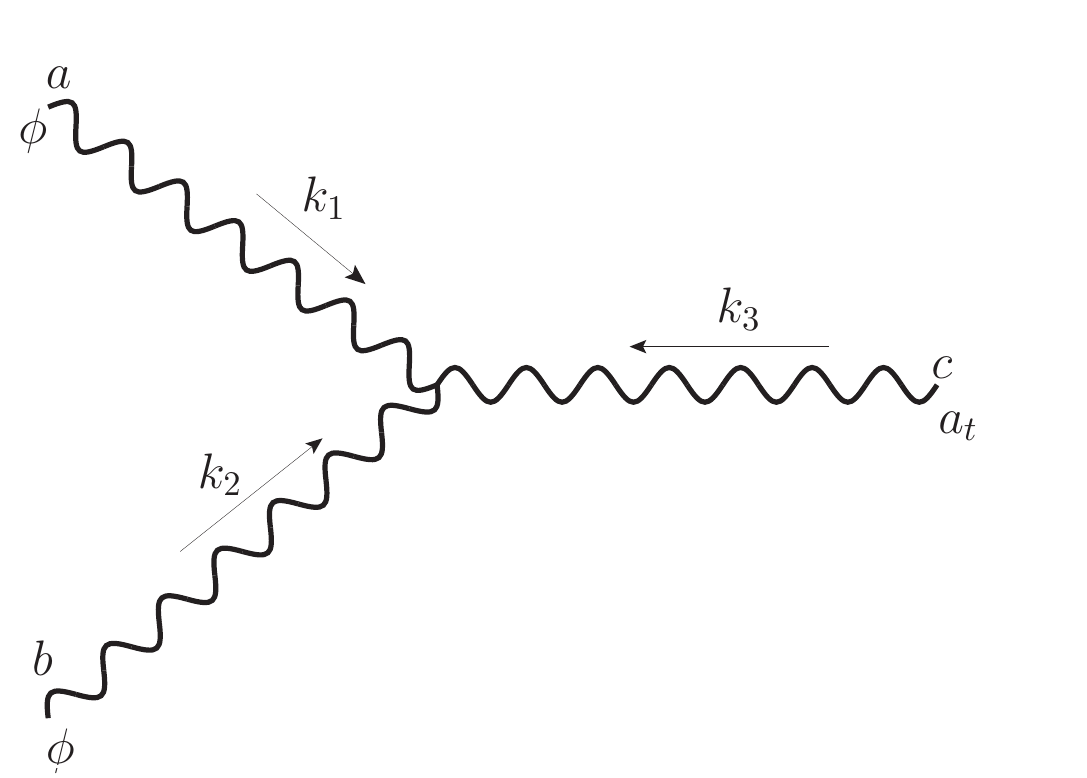}}
		\hspace{8pt}
		\subfigure[][$V_{3\,\phi a_t a_i}^{abc\,i}$]{
			\label{fig:ex3-b}
			\includegraphics[height=3.3cm]{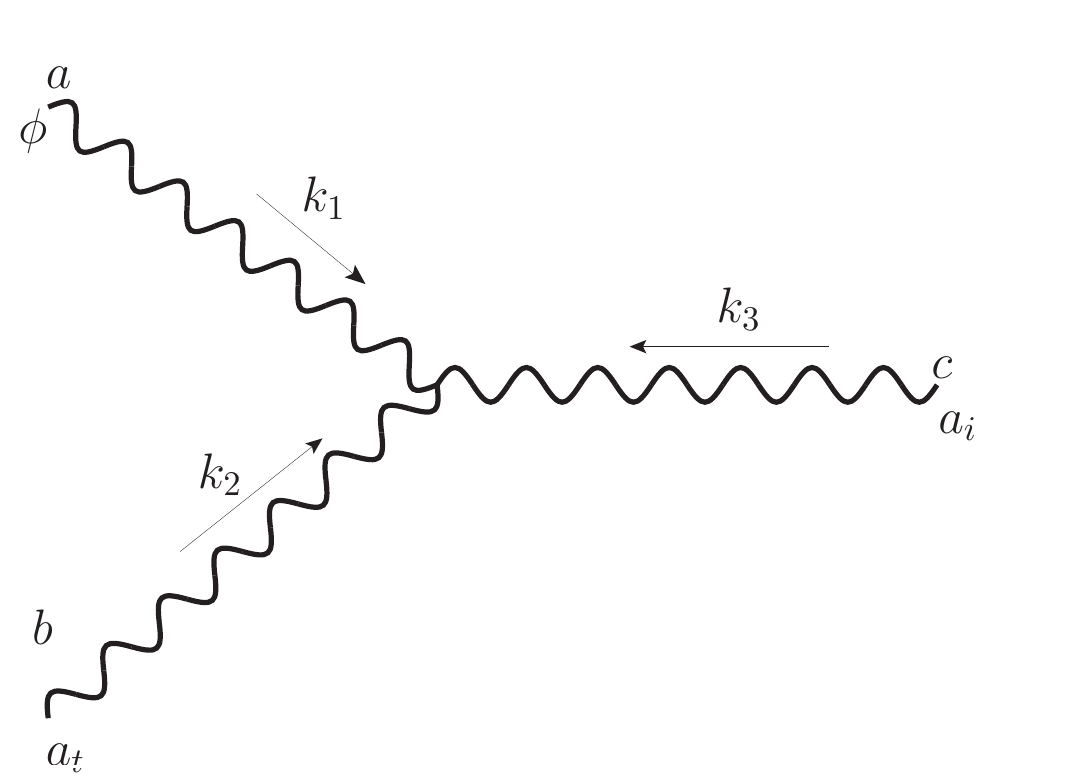}}
		\subfigure[][$V_{3\,\phi a_i a_j}^{abc\, ij}$]{
			\label{fig:ex3-c}
			\includegraphics[height=3.3cm]{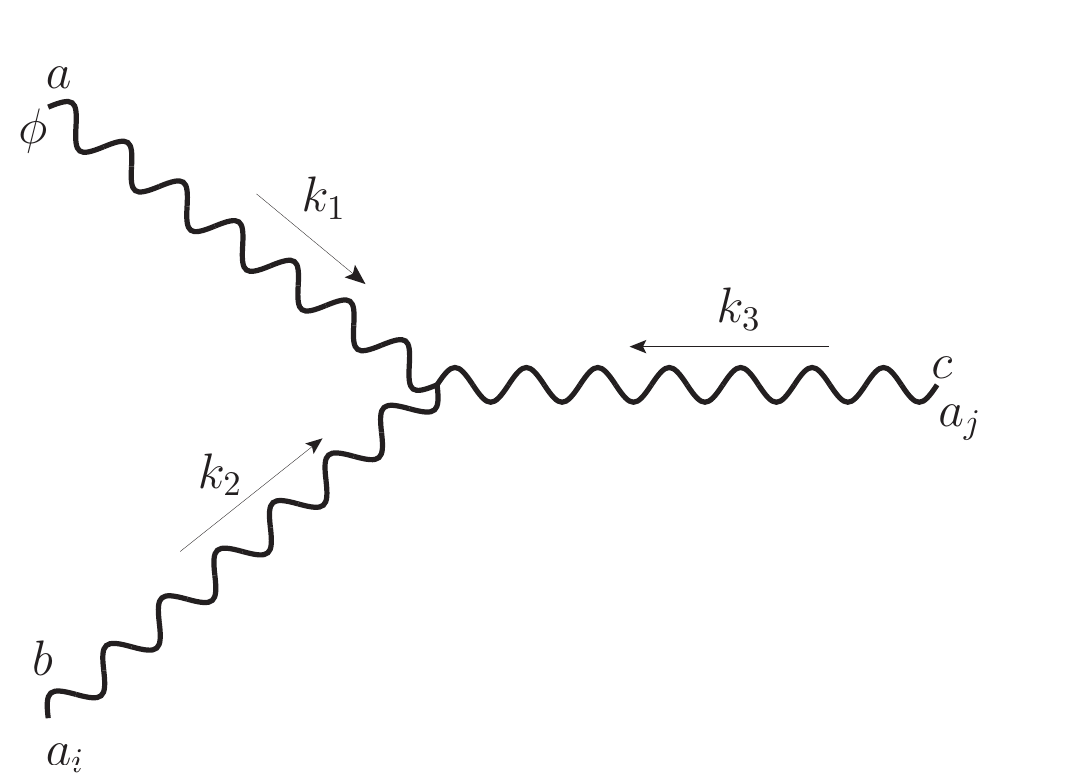}}\\
		\subfigure[][$V_{3\,a_i a_j a_k}^{abc\, ijk}$]{
			\label{fig:ex3-c}
			\includegraphics[height=3.3cm]{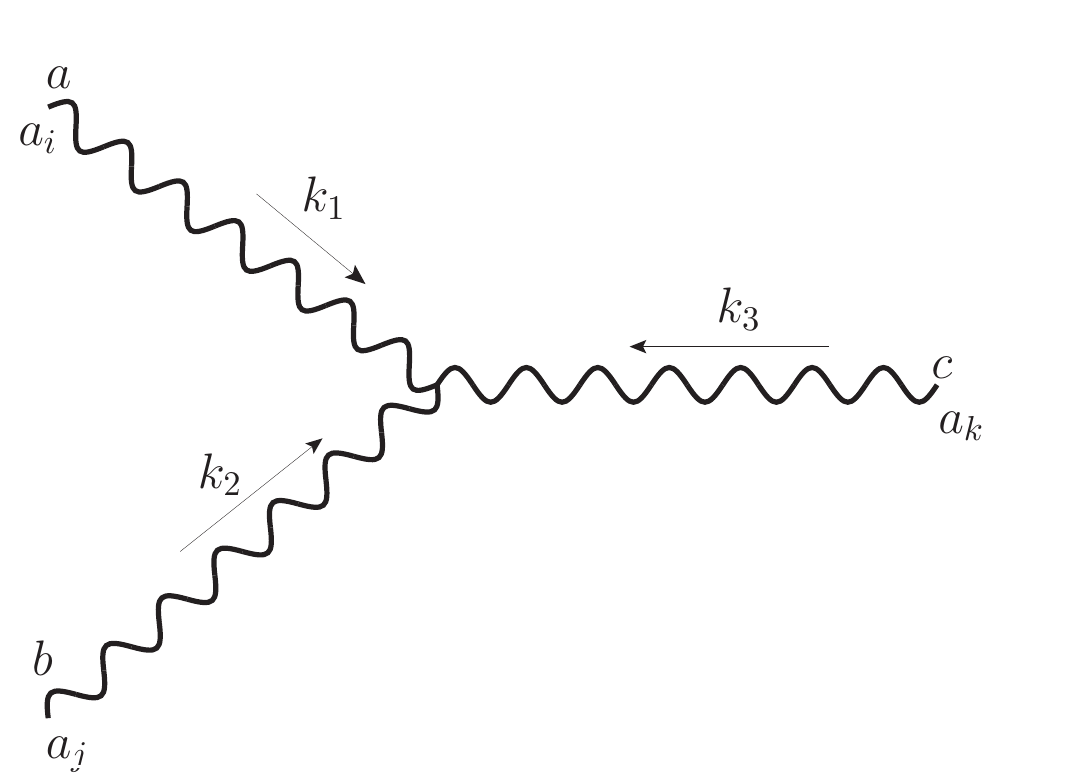}}
		\subfigure[][$V_{3\,\phi \bar{c} c}^{abc}$]{
			\label{fig:ex3-c}
			\includegraphics[height=3.3cm]{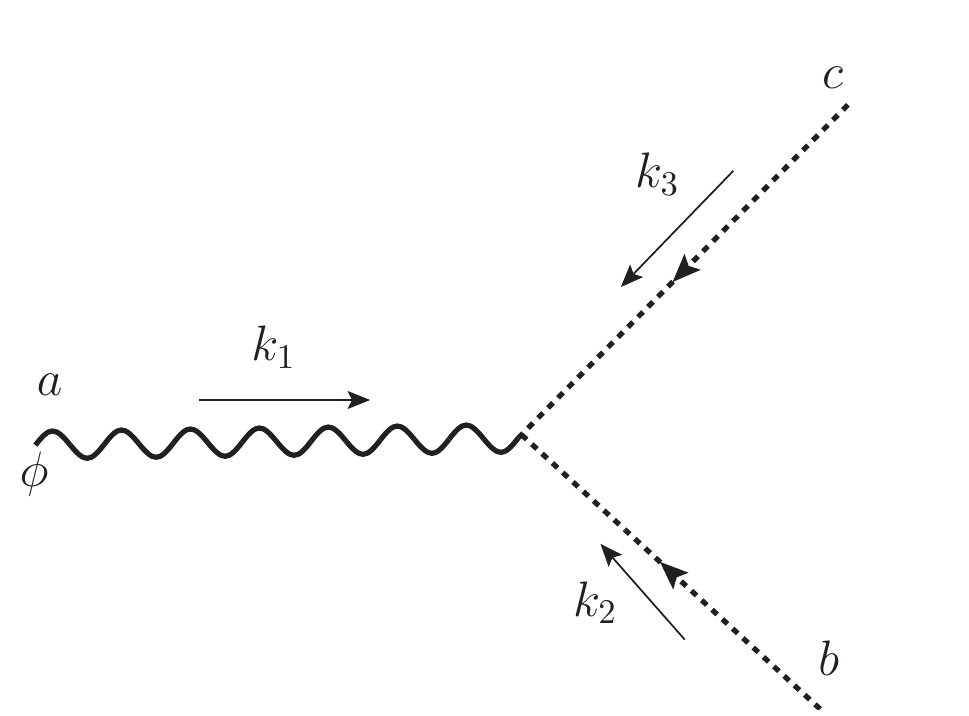}}
		\subfigure[][$V_{3\,a_i \bar{c} c}^{abc}$]{
			\label{fig:ex3-c}
			\includegraphics[height=3.3cm]{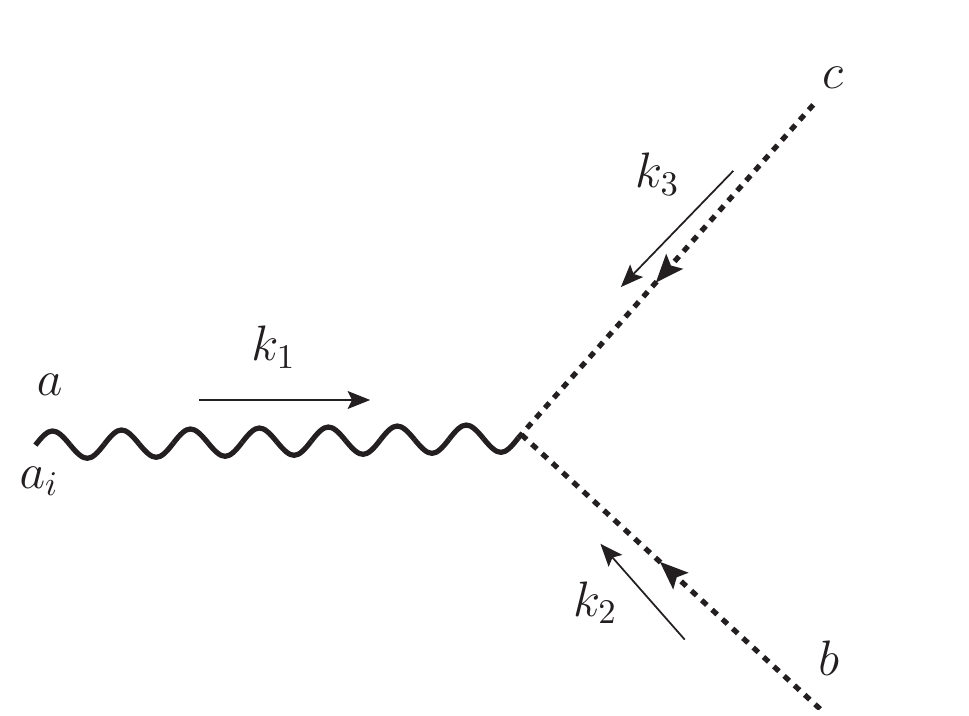}}
		\caption[Three-point Vertices.]{Three-point Vertices}
		\label{fig:ex3}
	\end{figure}
	
	\begin{figure}[h]
		\centering
		\subfigure[][$V_{4\,\phi\phi a_t a_t}^{bcde}$]{
			\label{fig:ex3-a}
			\includegraphics[height=3.3cm]{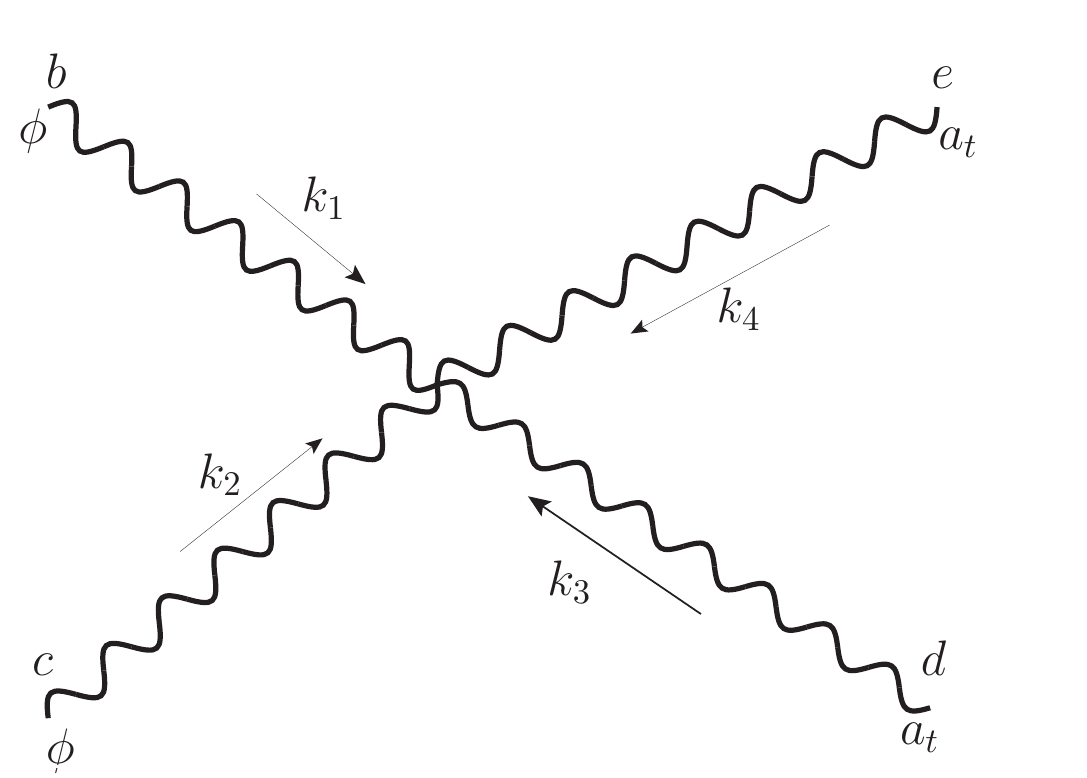}}
		\hspace{8pt}
		\subfigure[][$V_{4\,\phi a_t a_i a_j}^{bcde\,ij}$]{
			\label{fig:ex3-b}
			\includegraphics[height=3.3cm]{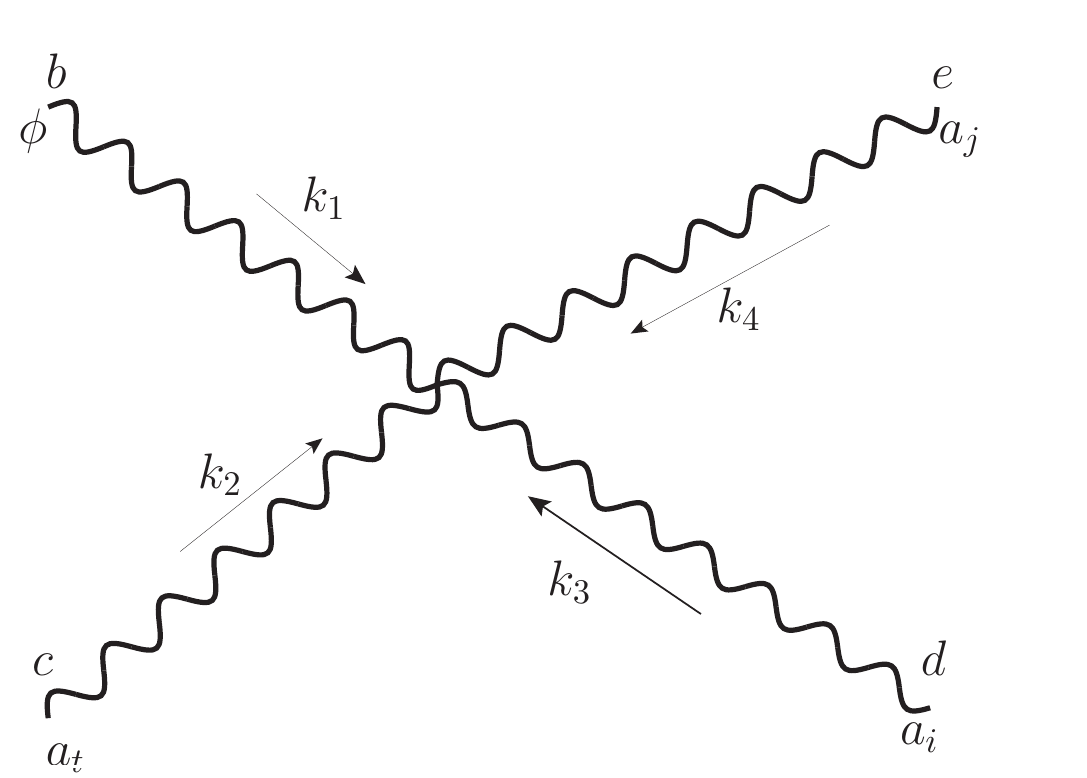}}
		\subfigure[][$V_{4\,a_i a_j a_k a_l}^{bcde\,ijkl}$]{
			\label{fig:ex3-c}
			\includegraphics[height=3.3cm]{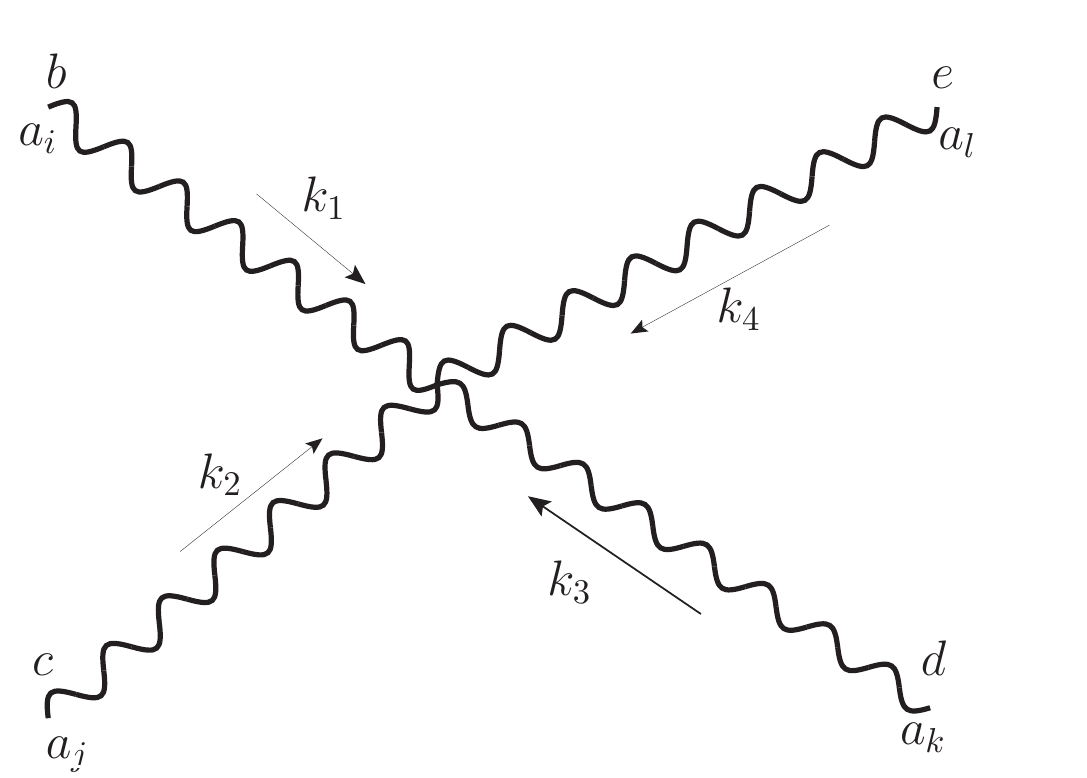}}
		\caption[Four-point Vertices.]{Four-point Vertices}
		\label{fig:ex3}
	\end{figure}
	
In future work, in order to make the theory contain non-trivial degrees of freedom, we would be coupling matter degrees of freedom to the above Lagrangian. The Feynman rules we have derived above would be useful for computing quantum mechanical processes in the non-abelian Galilean gauge theories, now coupled with matter. 

\bigskip

\newpage
	
	\section{ Conclusions and Future Directions}

\subsection*{Summary}
In this paper, we have investigated various properties of Galilean gauge theories. We have arrived at actions for abelian and non-Abelian Galilean QFTs by a process of null reduction from higher dimensional relativistic QFTs. We principally investigated the symmetry structures associated with these gauge theories and found that there are symmetry enhancements at various levels. 

For three dimensional theories (both abelian and non-abelian), the gauge theories exhibit invariance under the Schr{\"o}dinger group. The Schr\"odinger algebra admits an infinite extension giving the so-called Virasoro-Schr{\"o}dinger algebra. At the level of the action of these Galilean gauge theories, we found that the symmetries enhance from the finite version to a subset of the infinite algebra. At the level of the EOM, we found even more enhancements. Finally, in our analysis of the pre-symplectic structure, the symmetries in phase space exhibited the full Virasoro-Schro{\"o}dinger algebra, together with a state-dependent central extension. 

The story in four spacetime dimensions proceeded along similar lines, now with the Schr{\"o}dinger algebra replaced by the finite Galilean Conformal Algebra and the Virasoro-Schr{\"o}dinger algebra replaced by the infinite version of the GCA. 

Apart from this, as an aside, we also considered electro-magnetic duality in the $d=3$ abelian theory and found that unlike the $d=4$ case where the duality transformation exchanged electric and magnetic theories, the field equations of the theory we considered were invariant under the duality. This invariance was inherited from relativistic duality symmetric theory via null reduction. The $d=4$ case, even though also obtainable from null reductions, does not inherit any duality symmetry since the relativistic $d=5$ theory itself is not duality invariant. 

We concluded our analysis with a sneak-peek into the quantum-mechanical regime of these gauge theories and wrote down the propagators and vertices for the theories. The full-blown quantum analysis is left for future work. 
	
\subsection*{Looking ahead}

A lot of recent attention has been devoted to Carrollian theories, where the speed of light, instead of going to infinity like in Galilean theories, goes to zero \cite{LevyLeblond}. In these peculiar theories, the lightcones thus close up and a priori, these theories seem like non-sensical ones which would be of no use. Surprisingly however, these Carrollian theories have been shown to encompass a great deal of interesting physics. These Carrollian structures show up whenever one attempts to write down a QFT on a null surface \cite{Bagchi:2019clu, Ciambelli:2019lap} and hence is vitally important for understanding holographic duals to asymptotically flat spacetimes living on $\mathscr{I}^\pm$ \cite{Bagchi:2010zz, Bagchi:2012cy, Barnich:2012aw, Bagchi:2012xr, Barnich:2012xq, Bagchi:2014iea, Hartong:2015usd, Jiang:2017ecm, Ciambelli:2018wre}. These are also important for any generic null surface and hence crucial for theories defined on black hole event horizons \cite{Donnay:2019jiz}. It has been shown that 2d Carrollian CFTs (the symmetries of which are actually isomorphic to 2d Galilean CFTs \cite{Bagchi:2010zz}) replace usual relativistic CFTs on the worldsheet of strings in the tensionless or null limit \cite{Bagchi:2013bga, Duval:2014uva, Bagchi:2015nca, Bagchi:2019cay, Bagchi:2020ats, Bagchi:2021ban}. For higher dimensional objects like branes, higher dimensional Carrollian CFTs should be important. Of late, Carrollian theories have been found to be of importance in cosmological scenarios including dark energy \cite{deBoer:2021jej}, and in the theory of fractons  in condensed matter physics \cite{Bidussi:2021nmp}. Given such a wide array of applicability, field theories with Carrollian symmetry, especially Carrollian gauge theories are going to play a pivotal role going forward. Analysis of abelian and non-abelian Carrollian gauge theories have been carried out using EOM \cite{Bagchi:2019xfx}. It is however of greater importance to construct actions of quantum field theories with Carroll symmetry. Some work in this direction are \cite{Bagchi:2019clu, Banerjee:2020qjj, Hao:2021urq, Chen:2021xkw}. We would like to generalise these constructions to non-abelian theories and study their quantum mechanical structure.

Returning to the Galilean arena, we obviously wish to investigate the quantum mechanical structure of the gauge theories we have developed. This is work in progress. We wish to generalise the constructions in this paper to Supersymmetric QFTs. The process of null reduction would be useful in this case as well. We plan to return to these and other related questions in the near future.

\section*{Acknowledgments}
AB is partially supported by a Swarnajayanti fellowship of the Department of Science and Technology, India and by the following grants from the Science and Engineering Research Board: SB/SJF/2019-20/08, MTR/2017/000740, CGR/2020/002035. RB is supported by the grants SRG/2020/001037 and CRG/2020/002035 from the Science and Engineering Research Board. MI would like to thank Supratim Das Bakshi for useful discussions. AM would like to thank Alexander von Humboldt Foundation for a Humboldt Postdoctoral Research Fellowship which funded his tenure at AEI Potsdam.

	\newpage
	\appendix
	\section{Scale-Spin Representation of GCA}\label{app:representation}
	The GCA is given as
	\begin{eqnarray}\label{gcain}&&
	\big[L^{(n)},L^{(m)}\big]=(n-m)L^{(n+m)},~ \big[L^{(n)},M_{i}^{(m)}\big]=(n-m)M_{i}^{(n+m)},~ \big[M^{(n)}_{i},M^{(m)}_{j}\big]=0,\non\\&&
	\big[L^{(n)},J_{ij}\big]=0,~ \big[J_{ij},M^{(n)}_{k}\big]=M^{(n)}_{j}\delta_{ik}-M^{(n)}_{i}\delta_{jk}
	\end{eqnarray}
	where $L^{(-1,0,1)} \rightarrow (H,D,K),~M^{(-1,0,1)}_i \rightarrow (P_{i}, B_{i}, K_{i})$ respectively.
	We will now construct the representation theory for this algebra. We will begin by labelling the primary states \footnote{The primary states in the non-relativistic setting are defined as \begin{eqnarray}
		L^{n}|\varphi\rangle_{p}=M_{i}^{n}|\varphi\rangle_p=0, \quad \forall~ n>0.
		\end{eqnarray}} under dilatation ($L^{(0)}$) and rotation ($J_{ij}$) as \cite{Bagchi:2009ca,Bagchi:2014ysa,Bagchi:2017yvj}:
	\begin{eqnarray}
	L^{(0)}| \varphi \rangle_p =\Delta|\varphi\rangle_p,~ J_{ij}|\varphi\rangle_{p}=\Sigma_{ij}|\varphi\rangle_{p}.
	\end{eqnarray}
	where $\Delta$ is the scaling weight and $\Sigma_{ij}$ denotes the action of rotation in the particular representation of $SO(d - 1)$. The labeling of the states can be performed in this manner because both $D$ and $J_{ij}$ commutes \eqref{gcain} and we can use eigenvalues of both the operators simultaneously.
	We will write down the action of the finite part of GCA on the primary operators as
	\bes{}
	\begin{eqnarray}&&
	\big[L^{(-1)},\varphi(t,x)\big]=\partial_{t}\varphi(t,x),\quad \big[M_{i}^{(-1)},\varphi(t,x)\big]=-\partial_{i}\varphi(t,x)\\&&
	\big[J_{ij},\varphi(0,0)\big]=\Sigma_{ij}\varphi(0,0),\quad \big[L^{0},\varphi(0,0)\big]=\Delta\varphi(0,0)
	\end{eqnarray}\ees
	where we have used the state-operator correspondence (it relates the primary state and the vacuum as $|\varphi\rangle_p =\varphi(0,0)|0\rangle$) in the intermediate steps. Next step will be to calculate the action of $L^{(0)}$ and $J_{ij}$ on $\varphi$ at an arbitrary spacetime point $(t,x)$. For that, we will use
	\begin{eqnarray}
	\varphi(t,x)=U\varphi(0,0)U^{-1} \quad\text{with} \quad U=e^{tL^{(-1)}-x^{k}M_{k}^{(-1)}}
	\end{eqnarray}
	along with Baker-Campbell-Hausdrorff (BCH) formula and the commutation relations of GCA. The final result becomes
	\bes{}\label{dfs}
	\begin{eqnarray}&&
	\big[J_{ij},\varphi(t,x)\big]=\big(x_{i}\partial_{j}-x_{j}\partial_{i}\big)\varphi(t,x)+\Sigma_{ij}\varphi(t,x)\\&&
	\big[L^{(0)},\varphi({t,x})\big]=\big(t\partial_{t}+x^{k}\p_{k}+\Delta\big)\varphi(t,x)
	\end{eqnarray}\ees
	We will now look at the action of boost and SCT on the primaries. We know from \eqref{gcain} that boost and rotation generators do not commute. Its direct consequences will be that we cannot find the action of boost as simple as \eqref{dfs}. We have to think of another method to evaluate boost action on the primaries. One such way is to use the Jacobi identity given below:
	\begin{eqnarray}\label{jacob}
	\big[J_{ij},\big[B_{k},\varphi(0,0)\big]\big]=\big[B_{k},\Sigma_{ij}\varphi(0,0)\big]+\delta_{ki}\big[B_{j},\varphi(0,0)\big]-\delta_{jk}\big[B_{i},\varphi(0,0)\big]
	\end{eqnarray}
	By solving \eqref{jacob}, the action of boost on $\varphi(0,0)$ can be calculated for different theory. For our case where the field content is given by $(\phi \rightarrow \text{scalar field} ,\{a_t,a_i\}\rightarrow \text{gauge fields})$, we get
	\begin{eqnarray}\label{boot}
	\big[B_{k},\varphi(0,0)\big]=a\phi_{k}+s a_{k}+r \phi \delta_{ik}
	\end{eqnarray}
	where $\varphi(0,0)=\{\phi, a_t, a_i\}$. The values of constants $(a,s,r)$ are determined by demanding inputs from the dynamics.
	The action of boost on $\varphi(t,x)$ at finite spacetime points is given by
	\bea{}\label{boo}\big[B_{k},\varphi(t,x)\big]=-t\partial_{k}\varphi(t,x)
	+U\big[B_k,\varphi(0,0)\big]U^{-1}\eea
	Once we have \eqref{boot}, the action of SCT can be calculated easily. They are
	\bes{}\label{kki}
	\begin{eqnarray}&&
	\big[K,\varphi(t,x)\big]=\big(t^{2}\partial_{t}+2t x^{k}\p_{k}+2t \Delta\big)\varphi(t,x)
	-2 x^{k}U\big[B_k,\varphi(0,0)\big]U^{-1},\\&&
	\big[K_{k},\varphi(t,x)\big]=-t^2 \partial_{k}\varphi(t,x)
	+2 t U\big[B_k,\varphi(0,0)\big]U^{-1}.
	\end{eqnarray}\ees
	Similarly, the action of infinite dimensional generators ($L^{(n)}$ and $M^{(n)}_{i}$) on $\varphi(t,x)$ at arbitrary spacetime point can be found. They are given by
	\bes{}\label{gca rep1}
	\begin{eqnarray}&&
	\hspace{-1.5cm}\big[L^{(n)},\varphi(t,x)\big]=\big(t^{n+1}\partial_{t}+(n+1)t^{n}x^{k}\p_{k}+(n+1)t^{n}\Delta\big)\varphi(t,x)\non\\&&
	\hspace{2.7cm}-n(n+1)t^{n-1}x^{k}U\big[M_{k}^{0},\varphi(0,0)\big]U^{-1},\\&&
	\hspace{-1.5cm}\big[M^{(n)}_{k},\varphi(t,x)\big]=-t^{n+1}\partial_{k}\varphi(t,x)
	+(n+1)t^{n}U\big[M_{k}^{0},\varphi(0,0)\big]U^{-1}.
	\end{eqnarray}\ees

\newpage	
	
	\bibliographystyle{JHEP}
	\bibliography{GYM_draft}

\providecommand{\href}[2]{#2}\begingroup\raggedright\begin{thebibliography}{10}

\bibitem{Son:2008ye}
D.~T. Son, \emph{{Toward an AdS/cold atoms correspondence: A Geometric
  realization of the Schrodinger symmetry}},
  \href{https://doi.org/10.1103/PhysRevD.78.046003}{\emph{Phys. Rev. D}
  {\bfseries 78} (2008) 046003}
  [\href{https://arxiv.org/abs/0804.3972}{{\ttfamily 0804.3972}}].

\bibitem{Balasubramanian:2008dm}
K.~Balasubramanian and J.~McGreevy, \emph{{Gravity duals for non-relativistic
  CFTs}}, \href{https://doi.org/10.1103/PhysRevLett.101.061601}{\emph{Phys.
  Rev. Lett.} {\bfseries 101} (2008) 061601}
  [\href{https://arxiv.org/abs/0804.4053}{{\ttfamily 0804.4053}}].

\bibitem{Kachru:2008yh}
S.~Kachru, X.~Liu and M.~Mulligan, \emph{{Gravity duals of Lifshitz-like fixed
  points}}, \href{https://doi.org/10.1103/PhysRevD.78.106005}{\emph{Phys. Rev.
  D} {\bfseries 78} (2008) 106005}
  [\href{https://arxiv.org/abs/0808.1725}{{\ttfamily 0808.1725}}].

\bibitem{Bagchi:2009my}
A.~Bagchi and R.~Gopakumar, \emph{{Galilean Conformal Algebras and AdS/CFT}},
  \href{https://doi.org/10.1088/1126-6708/2009/07/037}{\emph{JHEP} {\bfseries
  07} (2009) 037} [\href{https://arxiv.org/abs/0902.1385}{{\ttfamily
  0902.1385}}].

\bibitem{2009}
S.~A. Hartnoll, \emph{Lectures on holographic methods for condensed matter
  physics},
  \href{https://doi.org/10.1088/0264-9381/26/22/224002}{\emph{Classical and
  Quantum Gravity} {\bfseries 26} (2009) 224002}.

\bibitem{LBLL}
M.~L. Bellac and J.-M. Levy-Leblond, \emph{{Galilean Electromagnetism}},
  {\emph{Nuovo Cimento.} {\bfseries 14B (1973)} }.

\bibitem{Bagchi:2014ysa}
A.~Bagchi, R.~Basu and A.~Mehra, \emph{{Galilean Conformal Electrodynamics}},
  \href{https://doi.org/10.1007/JHEP11(2014)061}{\emph{JHEP} {\bfseries 11}
  (2014) 061} [\href{https://arxiv.org/abs/1408.0810}{{\ttfamily 1408.0810}}].

\bibitem{Duval:2014uoa}
C.~Duval, G.~W. Gibbons, P.~A. Horvathy and P.~M. Zhang, \emph{{Carroll versus
  Newton and Galilei: two dual non-Einsteinian concepts of time}},
  \href{https://doi.org/10.1088/0264-9381/31/8/085016}{\emph{Class. Quant.
  Grav.} {\bfseries 31} (2014) 085016}
  [\href{https://arxiv.org/abs/1402.0657}{{\ttfamily 1402.0657}}].

\bibitem{Bagchi:2015qcw}
A.~Bagchi, R.~Basu, A.~Kakkar and A.~Mehra, \emph{{Galilean Yang-Mills
  Theory}}, \href{https://doi.org/10.1007/JHEP04(2016)051}{\emph{JHEP}
  {\bfseries 04} (2016) 051}
  [\href{https://arxiv.org/abs/1512.08375}{{\ttfamily 1512.08375}}].

\bibitem{VandenBleeken:2015rzu}
D.~Van~den Bleeken and C.~Yunus, \emph{{Newton-Cartan, Galileo-Maxwell and
  Kaluza-Klein}},
  \href{https://doi.org/10.1088/0264-9381/33/13/137002}{\emph{Class. Quant.
  Grav.} {\bfseries 33} (2016) 137002}
  [\href{https://arxiv.org/abs/1512.03799}{{\ttfamily 1512.03799}}].

\bibitem{Bergshoeff:2015sic}
E.~Bergshoeff, J.~Rosseel and T.~Zojer, \emph{{Non-relativistic fields from
  arbitrary contracting backgrounds}},
  \href{https://doi.org/10.1088/0264-9381/33/17/175010}{\emph{Class. Quant.
  Grav.} {\bfseries 33} (2016) 175010}
  [\href{https://arxiv.org/abs/1512.06064}{{\ttfamily 1512.06064}}].

\bibitem{Festuccia:2016caf}
G.~Festuccia, D.~Hansen, J.~Hartong and N.~A. Obers, \emph{{Symmetries and
  Couplings of Non-Relativistic Electrodynamics}},
  \href{https://doi.org/10.1007/JHEP11(2016)037}{\emph{JHEP} {\bfseries 11}
  (2016) 037} [\href{https://arxiv.org/abs/1607.01753}{{\ttfamily
  1607.01753}}].

\bibitem{Banerjee:2019axy}
K.~Banerjee, R.~Basu and A.~Mohan, \emph{{Uniqueness of Galilean Conformal
  Electrodynamics and its Dynamical Structure}},
  \href{https://doi.org/10.1007/JHEP11(2019)041}{\emph{JHEP} {\bfseries 11}
  (2019) 041} [\href{https://arxiv.org/abs/1909.11993}{{\ttfamily
  1909.11993}}].

\bibitem{Mehra:2021sfx}
A.~Mehra and Y.~Sanghavi, \emph{{Galilean electrodynamics: covariant
  formulation and Lagrangian}},
  \href{https://doi.org/10.1007/JHEP09(2021)078}{\emph{JHEP} {\bfseries 09}
  (2021) 078} [\href{https://arxiv.org/abs/2107.08525}{{\ttfamily
  2107.08525}}].

\bibitem{Duval:1993pe}
C.~Duval, \emph{{On Galileian isometries}},
  \href{https://doi.org/10.1088/0264-9381/10/11/006}{\emph{Class. Quant. Grav.}
  {\bfseries 10} (1993) 2217}
  [\href{https://arxiv.org/abs/0903.1641}{{\ttfamily 0903.1641}}].

\bibitem{Duval:2009vt}
C.~Duval and P.~A. Horvathy, \emph{{Non-relativistic conformal symmetries and
  Newton-Cartan structures}},
  \href{https://doi.org/10.1088/1751-8113/42/46/465206}{\emph{J. Phys. A}
  {\bfseries 42} (2009) 465206}
  [\href{https://arxiv.org/abs/0904.0531}{{\ttfamily 0904.0531}}].

\bibitem{Bagchi:2017yvj}
A.~Bagchi, J.~Chakrabortty and A.~Mehra, \emph{{Galilean Field Theories and
  Conformal Structure}},
  \href{https://doi.org/10.1007/JHEP04(2018)144}{\emph{JHEP} {\bfseries 04}
  (2018) 144} [\href{https://arxiv.org/abs/1712.05631}{{\ttfamily
  1712.05631}}].

\bibitem{Duval:1984cj}
C.~Duval, G.~Burdet, H.~P. Kunzle and M.~Perrin, \emph{{Bargmann Structures and
  Newton-cartan Theory}},
  \href{https://doi.org/10.1103/PhysRevD.31.1841}{\emph{Phys. Rev. D}
  {\bfseries 31} (1985) 1841}.

\bibitem{1991}
C.~Duval, G.~Gibbons and P.~Horváthy, \emph{Celestial mechanics, conformal
  structures, and gravitational waves},
  \href{https://doi.org/10.1103/physrevd.43.3907}{\emph{Physical Review D}
  {\bfseries 43} (1991) 3907–3922}.

\bibitem{1995}
B.~Julia and H.~Nicolai, \emph{Null-killing vector dimensional reduction and
  galilean geometrodynamics},
  \href{https://doi.org/10.1016/0550-3213(94)00584-2}{\emph{Nuclear Physics B}
  {\bfseries 439} (1995) 291–323}.

\bibitem{Santos:2004pq}
E.~S. Santos, M.~de~Montigny, F.~C. Khanna and A.~E. Santana, \emph{{Galilean
  covariant Lagrangian models}},
  \href{https://doi.org/10.1088/0305-4470/37/41/011}{\emph{J. Phys. A}
  {\bfseries 37} (2004) 9771}.

\bibitem{Chapman:2020vtn}
S.~Chapman, L.~Di~Pietro, K.~T. Grosvenor and Z.~Yan, \emph{{Renormalization of
  Galilean Electrodynamics}},
  \href{https://doi.org/10.1007/JHEP10(2020)195}{\emph{JHEP} {\bfseries 10}
  (2020) 195} [\href{https://arxiv.org/abs/2007.03033}{{\ttfamily
  2007.03033}}].

\bibitem{Taylor:2008tg}
M.~Taylor, \emph{{Non-relativistic holography}},
  \href{https://arxiv.org/abs/0812.0530}{{\ttfamily 0812.0530}}.

\bibitem{Hagen:1972pd}
C.~R. Hagen, \emph{{Scale and conformal transformations in galilean-covariant
  field theory}}, \href{https://doi.org/10.1103/PhysRevD.5.377}{\emph{Phys.
  Rev.} {\bfseries D5} (1972) 377}.

\bibitem{Niederer:1972zz}
U.~Niederer, \emph{{The maximal kinematical invariance group of the free
  Schrodinger equation.}},
  \href{https://doi.org/10.5169/seals-114417}{\emph{Helv. Phys. Acta}
  {\bfseries 45} (1972) 802}.

\bibitem{Henkel:1993sg}
M.~Henkel, \emph{{Schrodinger invariance in strongly anisotropic critical
  systems}}, \href{https://doi.org/10.1007/BF02186756}{\emph{J. Statist. Phys.}
  {\bfseries 75} (1994) 1023}
  [\href{https://arxiv.org/abs/hep-th/9310081}{{\ttfamily hep-th/9310081}}].

\bibitem{Bagchi:2009ca}
A.~Bagchi and I.~Mandal, \emph{{On Representations and Correlation Functions of
  Galilean Conformal Algebras}},
  \href{https://doi.org/10.1016/j.physletb.2009.04.030}{\emph{Phys. Lett.}
  {\bfseries B675} (2009) 393}
  [\href{https://arxiv.org/abs/0903.4524}{{\ttfamily 0903.4524}}].

\bibitem{Nishida:2007pj}
Y.~Nishida and D.~T. Son, \emph{{Nonrelativistic conformal field theories}},
  \href{https://doi.org/10.1103/PhysRevD.76.086004}{\emph{Phys. Rev. D}
  {\bfseries 76} (2007) 086004}
  [\href{https://arxiv.org/abs/0706.3746}{{\ttfamily 0706.3746}}].

\bibitem{Hellerman:2021qzz}
S.~Hellerman, D.~Orlando, V.~Pellizzani, S.~Reffert and I.~Swanson,
  \emph{{Nonrelativistic CFTs at Large Charge: Casimir Energy and Logarithmic
  Enhancements}},  \href{https://arxiv.org/abs/2111.12094}{{\ttfamily
  2111.12094}}.

\bibitem{Karananas:2021bqw}
G.~K. Karananas and A.~Monin, \emph{{More on the operator-state map in
  non-relativistic CFTs}},  \href{https://arxiv.org/abs/2109.03836}{{\ttfamily
  2109.03836}}.

\bibitem{Pellizzani:2021hzx}
V.~Pellizzani, \emph{{Operator spectrum of nonrelativistic CFTs at large
  charge}},  \href{https://arxiv.org/abs/2107.12127}{{\ttfamily 2107.12127}}.

\bibitem{Hellerman:2020eff}
S.~Hellerman and I.~Swanson, \emph{{Droplet-Edge Operators in Nonrelativistic
  Conformal Field Theories}},
  \href{https://arxiv.org/abs/2010.07967}{{\ttfamily 2010.07967}}.

\bibitem{Kravec:2018qnu}
S.~M. Kravec and S.~Pal, \emph{{Nonrelativistic Conformal Field Theories in the
  Large Charge Sector}},
  \href{https://doi.org/10.1007/JHEP02(2019)008}{\emph{JHEP} {\bfseries 02}
  (2019) 008} [\href{https://arxiv.org/abs/1809.08188}{{\ttfamily
  1809.08188}}].

\bibitem{Favrod:2018xov}
S.~Favrod, D.~Orlando and S.~Reffert, \emph{{The large-charge expansion for
  Schr\"odinger systems}},
  \href{https://doi.org/10.1007/JHEP12(2018)052}{\emph{JHEP} {\bfseries 12}
  (2018) 052} [\href{https://arxiv.org/abs/1809.06371}{{\ttfamily
  1809.06371}}].

\bibitem{Pal:2018idc}
S.~Pal, \emph{{Unitarity and universality in nonrelativistic conformal field
  theory}}, \href{https://doi.org/10.1103/PhysRevD.97.105031}{\emph{Phys. Rev.
  D} {\bfseries 97} (2018) 105031}
  [\href{https://arxiv.org/abs/1802.02262}{{\ttfamily 1802.02262}}].

\bibitem{Alishahiha:2009nm}
M.~Alishahiha, R.~Fareghbal, A.~E. Mosaffa and S.~Rouhani, \emph{{Asymptotic
  symmetry of geometries with Schrodinger isometry}},
  \href{https://doi.org/10.1016/j.physletb.2009.03.052}{\emph{Phys. Lett. B}
  {\bfseries 675} (2009) 133}
  [\href{https://arxiv.org/abs/0902.3916}{{\ttfamily 0902.3916}}].

\bibitem{Duval:2014lpa}
C.~Duval, G.~W. Gibbons and P.~A. Horvathy, \emph{{Conformal Carroll groups}},
  \href{https://doi.org/10.1088/1751-8113/47/33/335204}{\emph{J. Phys.}
  {\bfseries A47} (2014) 335204}
  [\href{https://arxiv.org/abs/1403.4213}{{\ttfamily 1403.4213}}].

\bibitem{Freedman:2012zz}
D.~Z. Freedman and A.~Van~Proeyen, \emph{{Supergravity}}. Cambridge Univ.
  Press, Cambridge, UK, 5, 2012.

\bibitem{Gomis:2020fui}
J.~Gomis, Z.~Yan and M.~Yu, \emph{{Nonrelativistic Open String and Yang-Mills
  Theory}}, \href{https://doi.org/10.1007/JHEP03(2021)269}{\emph{JHEP}
  {\bfseries 03} (2021) 269}
  [\href{https://arxiv.org/abs/2007.01886}{{\ttfamily 2007.01886}}].

\bibitem{Bagchi:2019clu}
A.~Bagchi, R.~Basu, A.~Mehra and P.~Nandi, \emph{{Field Theories on Null
  Manifolds}}, \href{https://doi.org/10.1007/JHEP02(2020)141}{\emph{JHEP}
  {\bfseries 02} (2020) 141}
  [\href{https://arxiv.org/abs/1912.09388}{{\ttfamily 1912.09388}}].

\bibitem{Beisert:2017pnr}
N.~Beisert, A.~Garus and M.~Rosso, \emph{{Yangian Symmetry and Integrability of
  Planar N=4 Supersymmetric Yang-Mills Theory}},
  \href{https://doi.org/10.1103/PhysRevLett.118.141603}{\emph{Phys. Rev. Lett.}
  {\bfseries 118} (2017) 141603}
  [\href{https://arxiv.org/abs/1701.09162}{{\ttfamily 1701.09162}}].

\bibitem{Beisert:2018zxs}
N.~Beisert, A.~Garus and M.~Rosso, \emph{{Yangian Symmetry for the Action of
  Planar $\mathcal N=$ 4 Super Yang-Mills and $\mathcal N=$ 6 Super
  Chern-Simons Theories}},
  \href{https://doi.org/10.1103/PhysRevD.98.046006}{\emph{Phys. Rev. D}
  {\bfseries 98} (2018) 046006}
  [\href{https://arxiv.org/abs/1803.06310}{{\ttfamily 1803.06310}}].

\bibitem{Banerjee:2020qjj}
K.~Banerjee, R.~Basu, A.~Mehra, A.~Mohan and A.~Sharma, \emph{{Interacting
  Conformal Carrollian Theories: Cues from Electrodynamics}},
  \href{https://doi.org/10.1103/PhysRevD.103.105001}{\emph{Phys. Rev. D}
  {\bfseries 103} (2021) 105001}
  [\href{https://arxiv.org/abs/2008.02829}{{\ttfamily 2008.02829}}].

\bibitem{Barnich:2011mi}
G.~Barnich and C.~Troessaert, \emph{{BMS charge algebra}},
  \href{https://doi.org/10.1007/JHEP12(2011)105}{\emph{JHEP} {\bfseries 12}
  (2011) 105} [\href{https://arxiv.org/abs/1106.0213}{{\ttfamily 1106.0213}}].

\bibitem{LevyLeblond}
L.~Leblond, \emph{{Une nouvelle limite non-relativiste du group de Poincaré}},
  {\emph{Annales Poincare Phys.Theor. 3 (1965) 1} }.

\bibitem{Ciambelli:2019lap}
L.~Ciambelli, R.~G. Leigh, C.~Marteau and P.~M. Petropoulos, \emph{{Carroll
  Structures, Null Geometry and Conformal Isometries}},
  \href{https://doi.org/10.1103/PhysRevD.100.046010}{\emph{Phys. Rev. D}
  {\bfseries 100} (2019) 046010}
  [\href{https://arxiv.org/abs/1905.02221}{{\ttfamily 1905.02221}}].

\bibitem{Bagchi:2010zz}
A.~Bagchi, \emph{{Correspondence between Asymptotically Flat Spacetimes and
  Nonrelativistic Conformal Field Theories}},
  \href{https://doi.org/10.1103/PhysRevLett.105.171601}{\emph{Phys. Rev. Lett.}
  {\bfseries 105} (2010) 171601}
  [\href{https://arxiv.org/abs/1006.3354}{{\ttfamily 1006.3354}}].

\bibitem{Bagchi:2012cy}
A.~Bagchi and R.~Fareghbal, \emph{{BMS/GCA Redux: Towards Flatspace Holography
  from Non-Relativistic Symmetries}},
  \href{https://doi.org/10.1007/JHEP10(2012)092}{\emph{JHEP} {\bfseries 10}
  (2012) 092} [\href{https://arxiv.org/abs/1203.5795}{{\ttfamily 1203.5795}}].

\bibitem{Barnich:2012aw}
G.~Barnich, A.~Gomberoff and H.~A. Gonzalez, \emph{{The Flat limit of three
  dimensional asymptotically anti-de Sitter spacetimes}},
  \href{https://doi.org/10.1103/PhysRevD.86.024020}{\emph{Phys. Rev. D}
  {\bfseries 86} (2012) 024020}
  [\href{https://arxiv.org/abs/1204.3288}{{\ttfamily 1204.3288}}].

\bibitem{Bagchi:2012xr}
A.~Bagchi, S.~Detournay, R.~Fareghbal and J.~Sim\'on, \emph{{Holography of 3D
  Flat Cosmological Horizons}},
  \href{https://doi.org/10.1103/PhysRevLett.110.141302}{\emph{Phys. Rev. Lett.}
  {\bfseries 110} (2013) 141302}
  [\href{https://arxiv.org/abs/1208.4372}{{\ttfamily 1208.4372}}].

\bibitem{Barnich:2012xq}
G.~Barnich, \emph{{Entropy of three-dimensional asymptotically flat
  cosmological solutions}},
  \href{https://doi.org/10.1007/JHEP10(2012)095}{\emph{JHEP} {\bfseries 10}
  (2012) 095} [\href{https://arxiv.org/abs/1208.4371}{{\ttfamily 1208.4371}}].

\bibitem{Bagchi:2014iea}
A.~Bagchi, R.~Basu, D.~Grumiller and M.~Riegler, \emph{{Entanglement entropy in
  Galilean conformal field theories and flat holography}},
  \href{https://doi.org/10.1103/PhysRevLett.114.111602}{\emph{Phys. Rev. Lett.}
  {\bfseries 114} (2015) 111602}
  [\href{https://arxiv.org/abs/1410.4089}{{\ttfamily 1410.4089}}].

\bibitem{Hartong:2015usd}
J.~Hartong, \emph{{Holographic Reconstruction of 3D Flat Space-Time}},
  \href{https://doi.org/10.1007/JHEP10(2016)104}{\emph{JHEP} {\bfseries 10}
  (2016) 104} [\href{https://arxiv.org/abs/1511.01387}{{\ttfamily
  1511.01387}}].

\bibitem{Jiang:2017ecm}
H.~Jiang, W.~Song and Q.~Wen, \emph{{Entanglement Entropy in Flat Holography}},
  \href{https://doi.org/10.1007/JHEP07(2017)142}{\emph{JHEP} {\bfseries 07}
  (2017) 142} [\href{https://arxiv.org/abs/1706.07552}{{\ttfamily
  1706.07552}}].

\bibitem{Ciambelli:2018wre}
L.~Ciambelli, C.~Marteau, A.~C. Petkou, P.~M. Petropoulos and K.~Siampos,
  \emph{{Flat holography and Carrollian fluids}},
  \href{https://doi.org/10.1007/JHEP07(2018)165}{\emph{JHEP} {\bfseries 07}
  (2018) 165} [\href{https://arxiv.org/abs/1802.06809}{{\ttfamily
  1802.06809}}].

\bibitem{Donnay:2019jiz}
L.~Donnay and C.~Marteau, \emph{{Carrollian Physics at the Black Hole
  Horizon}}, \href{https://doi.org/10.1088/1361-6382/ab2fd5}{\emph{Class.
  Quant. Grav.} {\bfseries 36} (2019) 165002}
  [\href{https://arxiv.org/abs/1903.09654}{{\ttfamily 1903.09654}}].

\bibitem{Bagchi:2013bga}
A.~Bagchi, \emph{{Tensionless Strings and Galilean Conformal Algebra}},
  \href{https://doi.org/10.1007/JHEP05(2013)141}{\emph{JHEP} {\bfseries 05}
  (2013) 141} [\href{https://arxiv.org/abs/1303.0291}{{\ttfamily 1303.0291}}].

\bibitem{Duval:2014uva}
C.~Duval, G.~W. Gibbons and P.~A. Horvathy, \emph{{Conformal Carroll groups and
  BMS symmetry}},
  \href{https://doi.org/10.1088/0264-9381/31/9/092001}{\emph{Class. Quant.
  Grav.} {\bfseries 31} (2014) 092001}
  [\href{https://arxiv.org/abs/1402.5894}{{\ttfamily 1402.5894}}].

\bibitem{Bagchi:2015nca}
A.~Bagchi, S.~Chakrabortty and P.~Parekh, \emph{{Tensionless Strings from
  Worldsheet Symmetries}},
  \href{https://doi.org/10.1007/JHEP01(2016)158}{\emph{JHEP} {\bfseries 01}
  (2016) 158} [\href{https://arxiv.org/abs/1507.04361}{{\ttfamily
  1507.04361}}].

\bibitem{Bagchi:2019cay}
A.~Bagchi, A.~Banerjee and P.~Parekh, \emph{{Tensionless Path from Closed to
  Open Strings}},
  \href{https://doi.org/10.1103/PhysRevLett.123.111601}{\emph{Phys. Rev. Lett.}
  {\bfseries 123} (2019) 111601}
  [\href{https://arxiv.org/abs/1905.11732}{{\ttfamily 1905.11732}}].

\bibitem{Bagchi:2020ats}
A.~Bagchi, A.~Banerjee and S.~Chakrabortty, \emph{{Rindler Physics on the
  String Worldsheet}},
  \href{https://doi.org/10.1103/PhysRevLett.126.031601}{\emph{Phys. Rev. Lett.}
  {\bfseries 126} (2021) 031601}
  [\href{https://arxiv.org/abs/2009.01408}{{\ttfamily 2009.01408}}].

\bibitem{Bagchi:2021ban}
A.~Bagchi, A.~Banerjee, S.~Chakrabortty and R.~Chatterjee, \emph{{A Rindler
  Road to Carrollian Worldsheets}},
  \href{https://arxiv.org/abs/2111.01172}{{\ttfamily 2111.01172}}.

\bibitem{deBoer:2021jej}
J.~de~Boer, J.~Hartong, N.~A. Obers, W.~Sybesma and S.~Vandoren, \emph{{Carroll
  symmetry, dark energy and inflation}},
  \href{https://arxiv.org/abs/2110.02319}{{\ttfamily 2110.02319}}.

\bibitem{Bidussi:2021nmp}
L.~Bidussi, J.~Hartong, E.~Have, J.~Musaeus and S.~Prohazka, \emph{{Fractons,
  dipole symmetries and curved spacetime}},
  \href{https://arxiv.org/abs/2111.03668}{{\ttfamily 2111.03668}}.

\bibitem{Bagchi:2019xfx}
A.~Bagchi, A.~Mehra and P.~Nandi, \emph{{Field Theories with Conformal
  Carrollian Symmetry}},
  \href{https://doi.org/10.1007/JHEP05(2019)108}{\emph{JHEP} {\bfseries 05}
  (2019) 108} [\href{https://arxiv.org/abs/1901.10147}{{\ttfamily
  1901.10147}}].

\bibitem{Hao:2021urq}
P.-x. Hao, W.~Song, X.~Xie and Y.~Zhong, \emph{{A BMS-invariant free scalar
  model}},  \href{https://arxiv.org/abs/2111.04701}{{\ttfamily 2111.04701}}.

\bibitem{Chen:2021xkw}
B.~Chen, R.~Liu and Y.-f. Zheng, \emph{{On Higher-dimensional Carrollian and
  Galilean Conformal Field Theories}},
  \href{https://arxiv.org/abs/2112.10514}{{\ttfamily 2112.10514}}.

\end{thebibliography}\endgroup

\end{document}